\documentclass[aip,reprint,twocolumn,showpacs,showkeys,amsmath,amssymb]{revtex4-1}

\usepackage{graphicx}
\usepackage{dcolumn}
\usepackage{bm}
\usepackage{color}
\usepackage{natbib}
\usepackage{hyperref} 
\usepackage{dsfont}
\usepackage{mathrsfs,amsmath} 
\usepackage{color}
\usepackage[normalem]{ulem}
\usepackage{soul}

\begin{document}

\title{Computing the linear viscoelastic properties of soft gels using an Optimally Windowed Chirp protocol.}

\author{Mehdi Bouzid*}
\email[]{mb1853@georgetown.edu}
\affiliation{Department  of  Physics,  Institute  for Soft  Matter  Synthesis and Metrology,
Georgetown  University, 37th and O Streets,  N.W., Washington,  D.C. 20057,  USA}

\author{Bavand Keshavarz*}
\email[]{bavand@mit.edu}
\thanks{*Contributed equally}
\affiliation{Department of Mechanical Engineering, Massachusetts Institute of Technology,
77 Massachusetts Avenue, Cambridge, Massachusetts 02139, USA}

\author{Michela Geri}
\email[]{mgeri@mit.edu}
\affiliation{Department of Mechanical Engineering, Massachusetts Institute of Technology,
77 Massachusetts Avenue, Cambridge, Massachusetts 02139, USA}

\author{Thibaut Divoux}
\email[]{divoux@crpp-bordeaux.cnrs.fr}
\affiliation{Centre de Recherche Paul Pascal, CNRS UMR~5031 - Universit\'e de Bordeaux, 
115 avenue Dr. Schweitzer, 33600 Pessac, France}
\affiliation{MultiScale Material Science for Energy and Environment, UMI~3466, CNRS-MIT, 
77 Massachusetts Avenue, Cambridge, Massachusetts 02139, USA}

\author{Emanuela Del Gado}
\email[]{ed610@georgetown.edu}
\affiliation{Department  of  Physics,  Institute  for Soft  Matter  Synthesis and Metrology,
Georgetown  University,  37th and O Streets,  N.W., Washington,  D.C. 20057,  USA}

\author{Gareth H. McKinley}
\email[]{gareth@mit.edu}
\affiliation{Department of Mechanical Engineering, Massachusetts Institute of Technology,
77 Massachusetts Avenue, Cambridge, Massachusetts 02139, USA}

\begin{abstract}
{We use molecular dynamics simulations of a model three-dimensional particulate gel, to investigate the linear viscoelastic response. The numerical simulations are combined with a novel test protocol (the optimally-windowed chirp or OWCh), in which a continuous exponentially-varying frequency sweep windowed by a tapered cosine function is applied. The mechanical response of the gel is then analyzed in the Fourier domain. We show that i) OWCh leads to an accurate computation of the full frequency spectrum at a rate significantly faster than with the traditional discrete frequency sweeps, and with a reasonably high signal-to-noise ratio, and ii) the bulk viscoelastic response of the microscopic model can be described in terms of a simple mesoscopic constitutive model. The simulated gel response is in fact well described by a mechanical model corresponding to a fractional Kelvin-Voigt model with a single Scott-Blair (or springpot) element and a spring in parallel. By varying the viscous damping and the particle mass used in the microscopic simulations over a wide range of values, we demonstrate the existence of a single master curve for the frequency dependence of the viscoelastic response of the gel that is fully predicted by the constitutive model. By developing a fast and robust protocol for evaluating the linear viscoelastic spectrum of these soft solids, we open the path towards novel multiscale insight into the rheological response for such complex materials.}
\end{abstract}


\maketitle 

\section{Introduction}
 Self-assembled soft solids with gel-like properties and a complex and hierarchical microstructure are commonly formed in colloidal suspensions, proteins and other biopolymers \cite{Mezzenga:2005,Lu:2008,Conrad:2010,Helgeson:2012,Gibaud:2013aa,Zhao:2014,Grindy:2015}. Their highly adaptive and tunable rheological response is of interest for novel technologies and smart material design, but distinguishing the role of different microstructural features over different lengthscales and timescales in order to fully understand and control the wide relaxation spectrum of these soft materials is extremely difficult. Recent advancements in experimental techniques have enabled accurate and efficient determination of the rheological response of soft materials across a broad range of linear and non-linear deformations\cite{Ewoldt:2008,Ewoldt:2010,Laurati:2011,Mao:2016a,Jaishankar:2013,Helal:2016,Aime:2016,Laurati:2017} and the combination of such approaches with imaging, ultrasound velocimetry or spectroscopy provides unique opportunities to bridge the gap between the macroscopic rheological behavior of a material and its micro- and even nano-scale structure/dynamics \cite{Cipelletti:2005,Mohraz:2005,Dibble:2008,Divoux:2010,Divoux:2011,Callaghan:2008,Manneville:2008,Chan:2013,Guo:2010,Perge:2014b,Tamborini:2014}. Nevertheless, constitutive models that capture the link between the microstructure and the mechanical response are still fundamentally lacking, and this limits quantitative interpretation of the rheological measurements. Computational models, in which the constitutive behavior emerges from a more microscopic and physically-grounded description of the gel structure and dynamics, can therefore play a crucial role in complementing experiments and theories. Recent studies have demonstrated that computational coarse-grained methods for soft matter can properly capture the structural and mechanical heterogeneities of soft gels, and help unravel and disentangle the microscopic processes underlying non-linear response, aging and hydrodynamic interactions in such materials \cite{Santos:2013,Park:2013,Colombo:2014,Varga:2015,Landrum:2016,Jamali:2017,Bouzid:2017,bouzid2018langmuir}. Combining such numerical approaches with advanced experimental techniques and appropriate quantitative constitutive models offers the potential to transform rheological studies of soft gels and advance our fundamental understanding of such versatile materials. 

Here we address two of the most formidable challenges in computational rheological studies of soft gels, i.e. ($i$) performing simulations that adequately probe the very broad width of their viscoelastic spectrum as well as ($ii$) overcoming the numerical fluctuations in the measured moduli, which requires large ensemble sizes and extensive computing time to obtain converged statistics. These concerns significantly limit the effectiveness and scope of computational studies. In the present study, we use the particle gel model introduced in ref.\cite{Colombo:2013}, which produces stable porous networks (even at low volume fractions) that feature extended relaxation spectra, microscopic dynamics and mechanics consistent with several observations in colloidal and protein gels\cite{Colombo:2013,Colombo:2014}. A typical snapshot of the model gel is shown in Fig.~\ref{Fig.1} (a), where only the interparticle links are shown for clarity. We perform a detailed numerical study of the model rheological response using a Non-Equilibrium-Molecular-Dynamics approach with overdamped equations of motion for the particles in athermal conditions (i.e. neglecting the thermal fluctuations). To help overcome the computational challenges mentioned above we use signal processing sequences adapted from radar \textit{chirp} sequences. Such an approach was first employed computationally by Visscher et al.\cite{Visscher1994} to evaluate the linear viscoelastic properties of ungelled Brownian dispersions. We extend this approach and reduce the numerical errors in the computed moduli by employing both amplitude- and frequency-modulated profiles similar to those used by bats and dolphins in echolocation\cite{Au2007}. In particular, we use a novel optimization scheme based on acoustical and optical signal processing algorithms that was recently developed for experimental measurements of linear viscoelasticity\cite{Geri:2018} and which is employed here for the first time in a numerical study. The resulting algorithm effectively reduces the time required to determine the viscoelastic spectrum by two orders of magnitude as well as eliminating ringing artifacts and fluctuations that otherwise can strongly affect such calculations\cite{Visscher1994}. 

These advancements allow us to directly and quantitatively evaluate the complex modulus $G^*(\omega)$ of the particulate gel over a wide range of frequencies $\omega$ and show that it can be compactly described by a fractional Kelvin-Voigt constitutive model (FKVM).
This model predicts a plateau in the elastic modulus at low frequencies (the equilibrium modulus $G_0$ of the gel), as well as a broad power-law dependence over a wide range of intermediate frequencies in the loss modulus. Such features reflect the very broad and self-similar spectrum of time- and length scales over which the microstructure can relax residual stresses in this type of materials. In fact, viscoelastic characteristics of this type have been observed experimentally in a wide range of different gelled and partially cross-linked systems (see for example refs.\cite{Chasset:1965,Chambon:1986,Winter:1997}) as well as in many biological materials\cite{Holt:2008,Nicolle:2010} and even capillary-bridged suspensions\cite{Koos:2011}. For polymeric gels and elastomers, molecular models have been developed\cite{Curro:1983,McKenna:1988} that integrate rubber elasticity theories of imperfectly-cross-linked networks with reptation dynamics of the dangling chains in order to describe quantitatively the power-law relaxation that is observed experimentally. However equivalent micromechanical models describing similar relaxation dynamics in attractive colloidal gels do not yet exist. Our comparison of the viscoelastic spectrum of the numerical gel and of the FKVM model is a first step toward constructing a constitutive model framework for soft particulate gels. The FKVM model is parameterized by only three material constants\cite{Jaishankar:2013} [see Fig.~\ref{Fig.1}~(b)] and we show below that it can provide a quantitative description of the viscoelastic properties of the attractive colloidal gels simulated numerically over 4.5 decades of dimensionless frequency (or Deborah number). Because of the computational efficacy of the Optimized Windowed Chirp algorithm we can thus rapidly evaluate the full, frequency-dependent complex modulus of a large number of simulated gels. The analysis provides scaling relationships that bring quantitative insight into how microscopic properties such as the viscous dissipation associated with damped particle motion and particle mass affect the macroscopic linear viscoelastic properties of the resulting gels. 

The remainder of this article is structured as follows. In section~\ref{numerical}, we outline the damped molecular simulation scheme and the gel preparation protocol. Section~\ref{OWCh} is dedicated to a detailed comparison between the Optimally Windowed Chirp method and traditional small amplitude oscillatory shear (SAOS) protocols which use discrete input frequencies to determine $G^*(\omega)$. The fractional Kelvin-Voigt model (FKVM) is introduced in section~\ref{KVM} and used to quantify the dependence of the gel complex modulus on the key parameters of the model in section \ref{dependency}. The study is concluded with a discussion in section~\ref{discussion}.

\section{Numerical model}
\label{numerical}

\subsection{Equations of motion}

\begin{figure}[t!]
\includegraphics[width=1\linewidth]{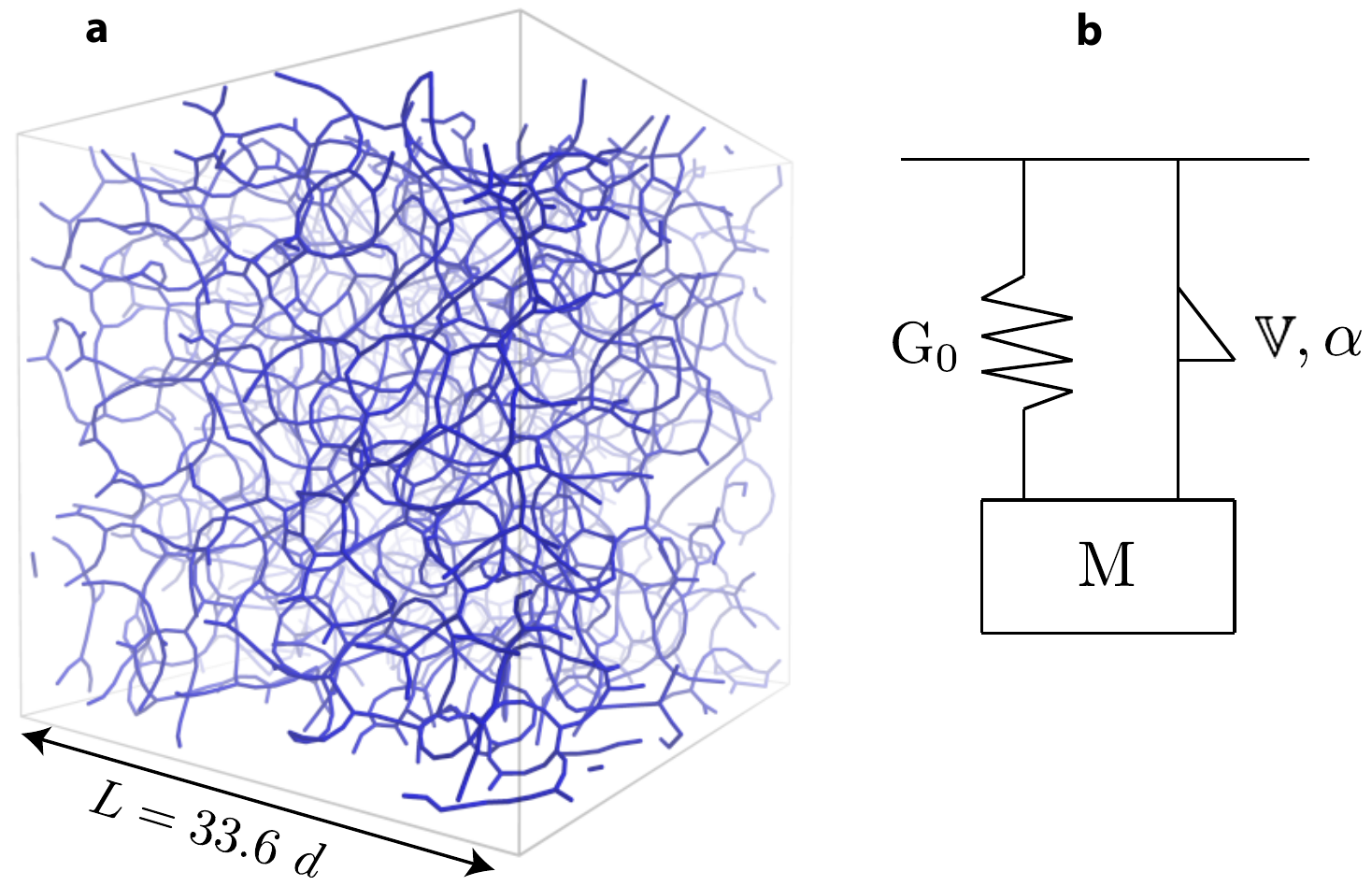}
\caption{(a) Snapshot of the colloidal gel network extracted from the simulation and formed at a number density $\rho=0.14$, which corresponds to a volume fraction $\phi\simeq 7.3\%$ and $N=5324$ particles. Each bond is represented by a segment, when the distance $d_{ij}$ between two particles $i$ and $j$ is $d_{ij}\leq 1.3d$. (b)~Schematic of the mechanical model of the gel. The model is composed of a mass $M$ connected to a spring of stiffness $G_0$ in parallel with a springpot, or fractional viscoelastic element, characterized by two parameters: a quasi-property $\mathds{V}$ (in \text{Pa$\cdot$s}$^\alpha$) and a dimensionless exponent $\alpha$.
\label{Fig.1}} 
\end{figure}

We perform molecular dynamics simulations of a model colloidal gel composed of $N$ particles each with a mass $m$ and diameter $d$ in a cubic simulation box of size $L$. The particles interact through a potential composed of two terms:
\begin{equation}
\mathcal{U}(\bold r_i,..., \bold r_N)=\epsilon \bigg{[}\sum_{i>j} \mathcal{U}_2\left(\frac{\bold r_{ij}}{d}\right)+\sum_i\sum_{j>k}^{j,k\neq i}\mathcal{U}_3\left(\frac{\bold r_{ij}}{d},\frac{\bold r_{ik}}{d}\right)\bigg]
\label{Pot}
\end{equation}
where $\bold{r}_{ij}=\bold{r}_j-\bold{r}_i$, with $\bold{r}_i$ denoting the position vector of the $i$-th particle, and $\epsilon$ the strength of the attraction that sets the energy scale. Typical values of $d$ and $\epsilon$ for colloidal particles range respectively from $d \simeq10$ to $100$~nm and from $\epsilon \simeq 10$ to $100$~$k_{B}\mathcal{T}$, with $k_B$ the Boltzmann constant and $\mathcal{T}$ the absolute temperature. The first contribution to $\mathcal{U}$ is a two-body potential \`a la  Lennard-Jones, $\mathcal{U}_2$, which consists of a repulsive core and a narrow attractive well that can be expressed in the following dimensionless form~:
\begin{equation}
\mathcal{U}_2(\bold{r})=A \left(\frac{a}{r^{18}}-\frac{1}{r^{16}}\right)
\end{equation}
where $r$ is the distance rescaled by the particle diameter $d$, while $a$ and $A$ are dimensionless parameters that control the width and the depth of the potential respectively. The second contribution to $\mathcal{U}$ is a three-body term $ \mathcal{U}_3$ that confers an angular rigidity to the inter-particle bonds, which prevents the formation of dense clusters. For two particles both bonded to a third one and whose relative position with respect to it are represented by the vectors $\mathbf{r}$ and $\mathbf{r}'$ (also rescaled by the particle diameter), it takes the following form:
\begin{equation}
\mathcal{U}_3(\bold{r},\bold{r'})=B \Lambda(\bold {r})\Lambda(\bold{r'})\exp\left[-\left(\frac{\bold{r}\cdot\bold{r'}}{rr'}-\cos \theta\right)^2 u^{-2}\right]
\end{equation}
where $B$, $\theta$ and $u$ are dimensionless parameters. The radial modulation $\Lambda(r)$ that controls the strength of the interaction reads:
\begin{equation}
\Lambda(r)= r^{-10}\left[1-(r/2)^{10}\right]^2 {\mathcal{H}}(2-r)
\end{equation}
where ${\mathcal{H}}$ denotes the Heaviside function, which ensures that $\mathcal{U}_3$ vanishes beyond the diameter of two particles. In conclusion, the potential energy (Eq.~\ref{Pot}) depends parametrically on five dimensionless quantities, which are fixed to the following values: $A=6.27$, $a=0.85$, $B=67.27$, $\theta=65^{\circ}$ and $u=0.3$. Tuning these parameters leads to a vast zoology of stable and porous microstructures. In the following, these values are chosen such that a disordered and thin percolating network starts to self-assemble for low particle volume fractions ($\phi \lesssim 0.1$), at $\epsilon=20k_B\mathcal{T}$, where $k_B$ is the Boltzmann constant and $\mathcal{T}$ is the absolute temperature. The self-assembly, the aging and the mechanical properties under external deformation of the resulting gel-like network structure have been studied extensively \cite{Colombo:2014b,Colombo:2013,Bouzid:2017,Colombo:2014} and exhibit several mechanical features consistent with the response measured in soft particulate gels in various experiments\cite{Derec:2003,Rajaram:2011,Sprakel:2011,Grenard:2014,Keshavarz:2017}.

\subsection{Initial configuration}
\label{initial}
The system is composed of $N$ particles in a cubic simulation box of size $L$ with periodic boundary conditions. The initial gel configuration is prepared with the protocol described in \cite{Colombo:2014}, which consists in starting from a gaseous configuration at $k_B\mathcal{T}/\epsilon=0.5$ and letting the gel self-assemble upon slow cooling down to $k_{B}\mathcal{T}/\epsilon =0.05$. The kinetic energy is then completely drawn from the system (down to $10^{-24}$) by means of a dissipative microscopic dynamics: 
\begin{equation}
m\frac{d^2\bold{r}_i}{dt^2}=-\nabla_{\bold{r_i}}\mathcal{U}-\eta_f\frac{d\bold{r}_i}{dt},
\end{equation}
where $ \eta_f$ is the damping coefficient associated with coupling of the particle motion to the surrounding fluid. The timestep $\delta t$ used for the numerical integration is $\delta t=0.005$. Distances are expressed in terms of the particle diameter $d$, masses are expressed in units of $m_{0}$, the energy in terms of the strength of the attraction $\epsilon$ and the time in the units of the characteristic timescale $\tau_0=\sqrt{{m_0}d^2/\epsilon}$. All data discussed here refer to a number particle density $ N/L^3 = 0.14$, which corresponds to an approximate solid volume fraction $\phi \simeq 7.3$, and to $N=19652$ and $L=52d$ (except to investigate the system size dependence where the number particle density has slightly been changed and set to $ N/L^3 = 0.16$). All simulations have been performed using a version of LAMMPS suitably modified by us \cite{plimpton1995fast}.
 
\subsection{Mechanical test and stress calculation}
To determine the gel mechanical viscoelastic properties, the particles are submitted to a continuous shear strain $\gamma(t)$ in the $xy$ plane according to the following equation: 
\begin{equation}
m\frac{d^2\bold{r}_i}{dt^2}=-\nabla_{\bold{r_i}}\mathcal{U}-\eta_f\left(\frac{d\bold{r}_i}{dt}-\dot\gamma(t)y_i\bold{e_x}\right)
\label{shear}
\end{equation}
The specific form of $\dot \gamma(t)$ will be introduced in the next section, and we use Lees-Edwards boundary conditions while applying the deformation \cite{Lees:1972}. 

It is worth noting that the equations of motion used here contain explicitly an inertial term (with mass $m$) for computational convenience, since this form allows for the use of effective and precise numerical integrators\cite{Frenkel:2001}. Nevertheless, the limit $m \rightarrow 0$ is the only one relevant to real colloidal gels in experiments, since in those systems the particle motion is completely overdamped and inertial effects are negligible. The timescales over which the particle motion is affected by inertia in our simulations are of the order $1\tau_{0} - 10\tau_{0}$ (for the values of $m$ and $\eta_{f}$ chosen). For a spherical colloidal (silica) particle of diameter $d \simeq 100nm$ and interaction strengths $\epsilon \simeq 10-20 k_{B} \mathcal{T}$, the inertial timescale $\tau_{0} \simeq 10^{-6}s$, i.e., it corresponds to timescales (and lengthscales, in terms of particle displacements) that are not accessible in typical rheometric experiments. Just for comparison, typical time scales associated with the viscous damping of the same particle when subjected to the same interactions in aqueous solution would be $\eta_{f}d^{2}/\epsilon \simeq5 \cdot 10^{-4}s$. As a consequence, for a finite value of $m$ the part of the viscoelastic spectra discussed in the following that is relevant to the experiments is only the one at low frequencies $\omega \ll \tau_0^{-1}$.

At the volume fraction used here, the gels are very soft due to the sparsely connected structure even in presence of relatively strong (with respect to $k_{B}\mathcal{T}$) interparticle interactions, hence we focus on the effect of the imposed deformation and neglect the role of thermal fluctuations in the structural rearrangements underlying the rheological response. Future work, building on the results obtained here, will be able to explore the changes of the linear viscoelastic spectrum due to the presence of thermal fluctuations that can assist in breaking network connections and redistributing stresses \cite{Bouzid:2017}.

We also note that for all frequencies considered, in absence of thermal fluctuations, there is no breakage of existing bonds, nor formation of new bonds. The deformation amplitude used in the oscillatory tests ($\gamma_{0} = 1\%$) is in the linear response regime, as extensively studied in \cite{Colombo:2014,bouzid2018langmuir}: in the absence of thermal fluctuations the linear response regime is substantially rate independent and can be estimated to extend up to strains $\gamma_{0} \approx 10\%$, while there is no bond broken or or newly formed up to strains  $\gamma_{0} \approx 50\%$.

The average state of stress of the gel is given by the virial stresses as $\sigma_{\alpha\beta}=-\frac{1}{L^3}\sum_{i}s_{\alpha\beta}^i$, where the Greek subscripts stand for the Cartesian components $x,y,z$ and $s_{\alpha\beta}^i$ represents the contribution to the stress tensor of all the interactions involving the particle $i$ \cite{irving1950statistical}. The latter contribution is calculated for each particle, by splitting  the contributions of the two-body and the three-body forces according to the following equation \cite{Colombo:2014}:
\begin{equation}
s^i_{\alpha\beta}=-\frac{1}{2}\sum_{n=1}^{N_2}(r_\alpha^iF_\beta^i+r_\alpha^{\prime}F_\beta^{\prime})+\frac{1}{3}\sum_{n=1}^{N_3}(r_\alpha^iF_\beta^i+r_\alpha^{\prime}F_\beta^{\prime}+r_\alpha^{\prime\prime}F_\beta^{\prime\prime})
\label{vstress}
\end{equation}
The first term denotes the contribution of the two-body interaction, where the sum runs over all the $N_2$ pairs of interactions that involve the particle $i$. The couples $(r^i,F^i)$ and $(r^{\prime},F^{\prime})$ denote respectively the position and the forces on the two interacting particles. In the same way, the second term indicates the three-body interactions involving the particle $i$ and two neighbors denoted by the prime and double prime quantities. 

\section{Linear frequency response}

For each gel self-assembled following the procedure described in Section \ref{initial}, we investigate its linear viscoelastic properties in the athermal limit (i.e. $k_{B}\mathcal{T}/\epsilon \simeq 0$). Similar to experiments, the viscoelastic response of the gel at a discrete frequency $\omega_i$ can also be measured in simulations by imposing an oscillatory shear strain $\gamma(t)=\gamma_0\sin(\omega_i t)$ and monitoring the corresponding response through the shear component $\sigma(t)$ of the stress tensor over a finite time $T$. Assuming a linear response regime, the elastic and loss moduli, $G'$ and $G''$ respectively, are calculated as
\begin{equation}
\begin{split}
&G'(\omega_i)=\mathcal{R}e\left\{\frac{\tilde{\sigma}(\omega_i)}{\tilde{\gamma}(\omega_i)}\right\}\\
&G''(\omega_i)=\mathcal{I}m\left\{\frac{\tilde{\sigma}(\omega_i)}{\tilde{\gamma}(\omega_i)}\right\}
\label{FFT}
\end{split}
\end{equation}
where $\tilde{\sigma}$ and $\tilde{\gamma}$ are the Fourier transforms of the stress and strain signals respectively \cite{Macosko:1994}. The whole viscoelastic spectrum is then reconstructed by performing a discrete series of tests at various frequencies, also known  as ``frequency sweep''. The finite duration of the input signal leads to the appearance of artificial components in the frequency spetrum, also referred to as ``spectral leakage'', which limit the accuracy of the values of $G'$ and $G''$ obtained. For a periodic signal \cite{Pintelon2012}, the spectral leakage can be reduced by choosing $T=n(2\pi/\omega_i)$, with integer values of $n \geq 1$, which requires a minimum signal length of $T=2\pi/\omega_i$. This  requires very long tests especially for measurements at low frequencies. In Brownian dynamics simulations, the signal/noise ratio also decreases at low frequencies because of the low values of the dimensionless strain rate, or bead-P\'eclet number, often necessitating the use of variance reduction techniques. Minimizing the measurement time is essential to reduce the computational effort, and is also of great importance, for example, in experiments studying rapidly gelling systems \cite{Winter:1986,Mours1994,Winter:1997}. To overcome these issues, a compact signal in the time domain, that spans over a broad range in the frequency domain, is desired. Holly et al. \cite{Holly1988} suggested a ``multi-wave" technique based on applying a waveform that is a linear superposition of a fundamental frequency and a few of its corresponding harmonics. This technique indeed enabled researchers to measure the viscoelastic properties of different gels at several frequencies with an experimental duration that is much shorter than with the discrete frequency approach \cite{Tang2009,In1993,Ross-Murphy1994,Pogodina1999,Schwittay1995,Chiou1996}. However, in such a multi-wave method, the amplitude of the multi-wave input signal is not constant and each modes contribution to the total strain can combine additively, exceeding the linear limit of the material, or combine subtractively, and thus fall below the sensitivity of the instrument sensor.

Inspired by studies of signal design for radar and acoustic applications\cite{Klauder1960,Farina2000,Fausti2000}, Ghiringhelli et al.\cite{Ghiringhelli:2012} more recently used a \textit{chirp} signal for studying the rheology of alginate gels. The signal consists of an oscillating trigonometric signal with a phase angle that is exponentially increasing with time over a predetermined range of selected frequencies. They showed that with this compact signal one can rapidly determine the linear viscoelastic behavior of the material over the specified range of frequencies. The compactness of this measurement technique, coined ``Optimal Fourier Rheometry" (OFR) \cite{Ghiringhelli:2012}, inspired Curtis and coworkers \cite{Curtis:2015} to use OFR to study the behavior of a rapidly gelling collagen system. These experimental studies demonstrated that chirp signals can indeed be successfully used to obtain the mechanical spectrum of a time-evolving gel over almost a decade in frequency throughout the gelation process. Earlier numerical studies by Heyes and coworkers\cite{Visscher1994,Heyes1994,Heyes1994a} used similar types of chirp signals in Brownian dynamics simulations of hard spherical dispersions to reduce the computational time required to evaluate the viscoelastic spectrum. However, their calculations of the resulting viscoelastic moduli were affected by spectral leakage, featuring fluctuations that could not be entirely eliminated even after post-processing the signal through a short-time--Fourier-transform using a Gaussian filter \cite{Visscher1994}.

In fact, the short duration of the input signal in the OFR technique, despite being a key feature, also strongly affects the precision of the measurements. Spectral leakage at frequencies beyond the minimum and maximum imposed values and set by the signal duration and the sampling frequency respectively, leads to ringing artifacts in the frequency spectrum. The ringing is also known as the Gibbs phenomenon in signal processing \cite{Blackman1958} and results in artificial fluctuations in the computed values of the frequency-dependent storage and loss moduli obtained from the chirp input signals\cite{Visscher1994}. Studies of spectral analysis and signal design have shown that by using windowing functions, which modulate the amplitude of the input signal, one can significantly reduce the leakage error in the resulting Fourier analysis \cite{Harris1978,Pintelon2012}. Recently, Geri et al.\cite{Geri:2018} used amplitude modulation of exponential chirp signals to perform high accuracy rheological measurements. By enveloping the exponentially-varying chirp signal with a symmetrically tapered window (known as a Tukey window \cite{Tukey1967}), Geri and co-workers demonstrated on several materials, including a semi-dilute entangled polymer solution and a time-evolving cross-linked biopolymer gel, that a significant reduction in the spectral leakage error can be achieved, improving the quantitative determination of the elastic and loss moduli.
The present study builds upon this Optimally Windowed Chirp (OWCh) method by using an exponential chirp signal whose amplitude is a tapered Tukey window as the strain input to the numerically-simulated gels. The tapering ratio used here is described below and is similar to the optimum value determined by Geri et al.\cite{Geri:2018}.

\begin{figure}[t!]
\includegraphics[width=\linewidth]{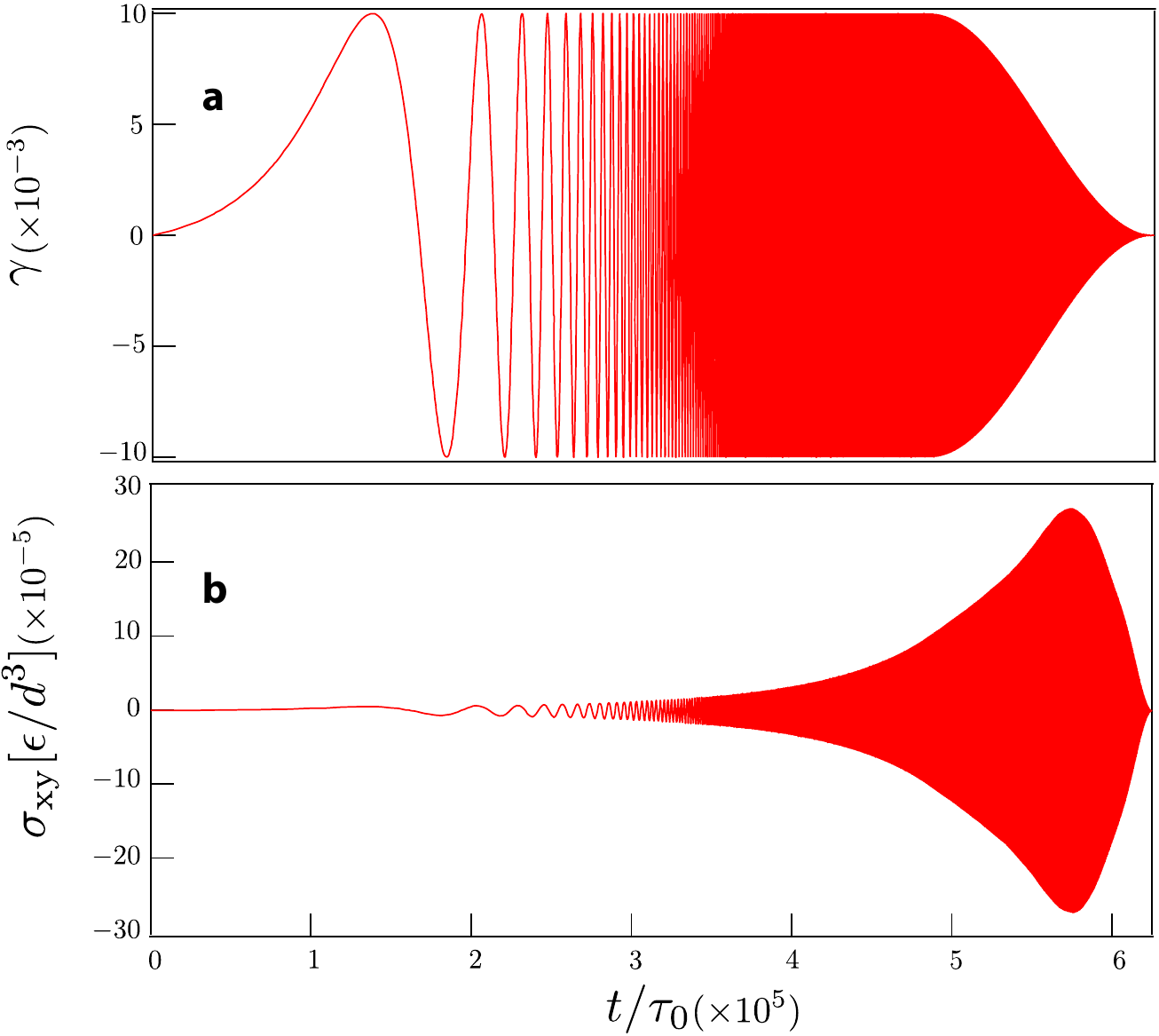}
\caption{\label{fig.1} (a) The input strain signal that is applied to the gel is an exponential chirp (Eq. \ref{expchirp}) with $\omega_1=10^{-4}~\tau_0^{-1}$, $\omega_2= 18~\tau_0^{-1}$,  $T=62510~\tau_0$, and a corresponding time-bandwidth product $TB\simeq1.79\times10^5$. The amplitude envelope is a Tukey window with a tapering parameter $b=0.45$. (b)~Resulting shear stress $\sigma_{xy}$ measured as a function of time. 
\label{Fig.2}} 
\vspace{-0.5cm}
\end{figure}

\subsection{Optimally windowed chirp rheometry}
\label{OWCh}

In this subsection, we compare quantitatively the results of traditional small amplitude oscillatory shear flow (SAOS) at discrete input frequencies with that of the Optimally Windowed Chirp (OWCh) method for determining the linear viscoelastic properties of the gel we simulate numerically. We first consider the traditional discrete frequency sweep consisting of an imposed oscillatory shear applied to the system following Eq.~\ref{shear}, where the shear strain is modulated periodically according to $\gamma(t)=\gamma_0 \sin(\omega_i t)$. The strain amplitude is fixed to $\gamma_0=0.01$ which is known to be in the linear response regime for the model gel \cite{Colombo:2014} and the frequency $\omega_i$ is changed in discrete steps to explore the viscoelastic spectrum over five orders in magnitude of frequency. For each frequency $\omega_i$, the signal duration is chosen to be an integer multiple of the signal period to avoid spectral leakage in the subsequent Fourier analysis. 
We contrast this classical approach with the OWCh scheme in which we apply an exponential chirp signal for the input strain [Fig.~\ref{Fig.2}(a)] that reads:
\begin{equation}\label{expchirp}
\gamma(t) =W(t;b)\gamma_0\sin\left[(L\omega_1)(e^{t/L}-1)\right]\\
\end{equation}
where $\gamma_0$ is the strain amplitude and $L=T/\ln(\omega_2/\omega_1)$ is the characteristic time over which the phase of the signal exponentially grows from the initial frequency $\omega_1$ to the final frequency $\omega_2$ within the duration of the signal $T$.  The window function $W(t;b)$ sets the shape of the amplitude envelope. It is taken here to be an asymmetric Tukey window with the following form:
\begin{equation}\label{window}
W(t;b)=\begin{Bmatrix}
1  \:\:\: \:\:\:\:\:\:\:\:\:\:\:\:\:\:\:\:\:\:\:\:\:\:\:\:\:\:\:\text{     if }0\le t/T\leq (1-b/2)
\\ \\\cos^2\left[{\frac{\pi}{b}}\left(\frac{t}{T}-1+\frac{b}{2}\right)\right]\:\text{ if } (1-b/2)\leq t/T
\end{Bmatrix}
\end{equation}
where the tapering parameter $b$ is the duration of the falling part of the envelope normalized by $T$. The duration of the signal $T$ is kept close to the period of the lowest imposed frequency, so that $T\sim \mathcal{O} (2\pi/\omega_1)$.\\
A dimensionless parameter that characterizes the leakage behavior of a chirp signal in the spectral domain is the time-bandwidth product $TB\equiv T(\omega_2-\omega_1)/2\pi$. High values of the time-bandwidth product reduce the artifacts associated with the ringing of the data\cite{Klauder1960}. For a fixed frequency range, this can be achieved by increasing the signal length $T$. However, in rapidly-gelling systems a low value of $T$ is also a vital feature. As shown by Geri et al. \cite{Geri:2018}, for a fixed value of $TB$ (and in the absence of noise in the output signal) a range of $0.1\leq b \leq 1$ optimizes the signal envelope and significantly reduces the measurement errors compared to the non-windowed signals ($b=0$) (see Fig.~\ref{Fig.9} in the appendix). By monitoring the shear stress response of the material $\sigma_{xy}(t)$ over time [Fig.~\ref{Fig.2}(b)], we can extract the viscoelastic moduli using the same Fourier transform protocol displayed in Eq.~\ref{FFT}.

\begin{figure}[t!]
\includegraphics[width=1\linewidth]{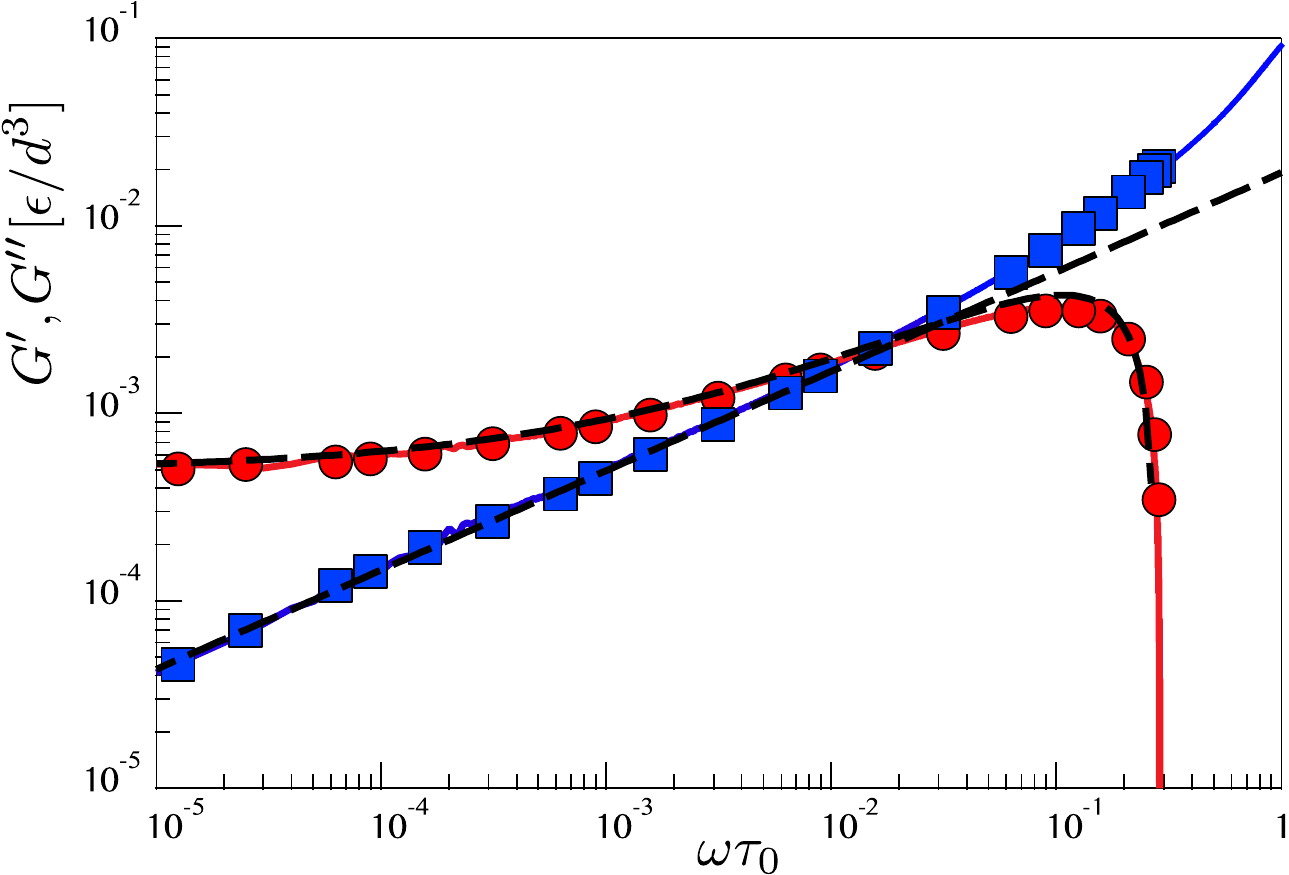}
\caption{Storage and loss moduli, $G'$ (\textcolor{blue}{$\bullet$}) and $G''$ (\textcolor{blue}{$\blacksquare$}) as a function of the frequency $\omega$. The symbols are calculated using the classical discrete frequency sweep, while the continuous red and blue lines are measured using the OWCh protocol. The black dashed line corresponds to the inertialess limit and the best fit of the data by the fractional model introduced in Fig.~\ref{Fig.1}(b), with $G_0=5.10^{-4}\epsilon/d^3$ , $\mathds{V}=0.025$ and $\alpha=0.53$. 
\label{Fig.3}} 
\end{figure}

The results obtained with both the discrete SAOS and the OWCh methods are displayed in Fig.~\ref{Fig.3} as discrete symbols and continuous lines respectively. Both methods lead to the same quantitative results. The gel behaves as a soft solid at low frequency with an asymptotic equilibrium modulus $G_0$ as $\omega\tau_0\rightarrow 0$. As the dimensionless oscillatory frequency increases, both moduli display a power law increase up to a crossover point beyond which the elastic modulus becomes smaller than the viscous modulus. At higher frequencies, the elastic modulus exhibits a maximum $G'_{\rm{max}}$ at a characteristic frequency we denote $\omega_c$, before experiencing an abrupt cutoff at larger frequencies. By contrast, the viscous modulus shows a power-law response over the entire range of frequency. The maximum in $G'(\omega)$ can be understood because each particle in our simulations is, as a matter of fact, a non-linear harmonic oscillator with mass $m$ and subjected to a viscous damping $\eta_{f}$ (see eq.\ref{shear}). The presence of inertia introduces a resonance in the spectrum, which manifests itself in a change in sign in the in-phase contribution of the response (i.e., the storage modulus $G'(\omega)$) at high frequencies. Inertia induced resonances in rheological measurements are discussed at more length in \cite{Walters1975a}.

\begin{figure}[!t]
\includegraphics[width=1\linewidth]{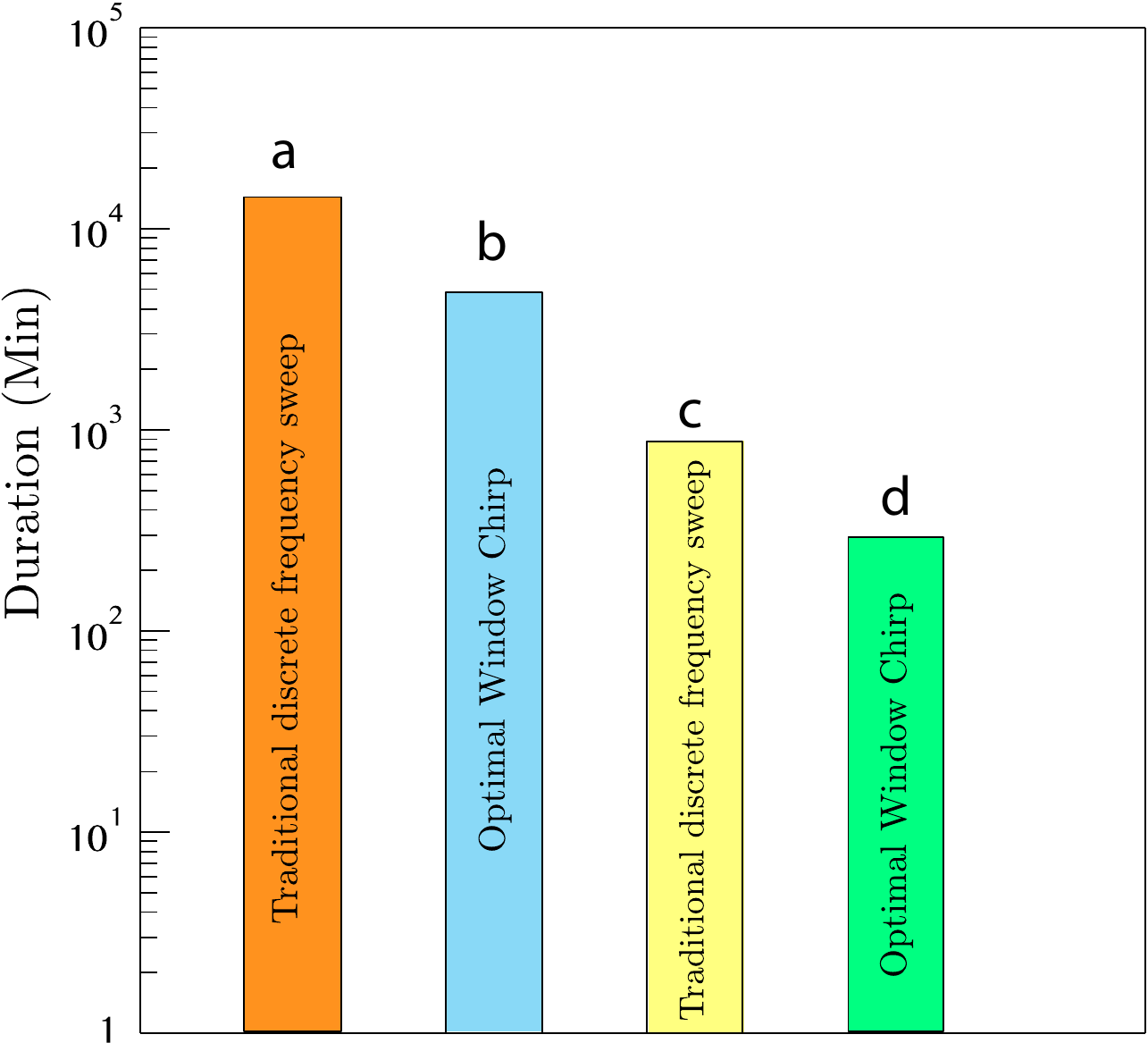}
\caption{Total duration of the simulations to measure numerically the frequency dependence of the viscoelastic moduli over a frequency range $\omega\in[10^{-4}\tau_0^{-1},\tau_0^{-1}]$: (a) Using the traditional discrete frequency sweep with 20 discrete frequency points. Each data point corresponds to an independent simulation performed with a single-core CPU and the total duration is calculated by summing the duration of the 20 runs. (b) Using the OWCh method with a single-core CPU. (c) Using the traditional discrete frequency sweep in a parallelized simulation distributed over a multi-core processor with 20 cores, for each frequency. (d) Using the OWCh method in a parallelized simulation as in (c). The simulation performance are improved by a factor of $\sim 3$ in (b) a factor of $\sim 16$ in (c) and by a factor of $\sim 50$ in (d).
\label{Fig.4}} 
\end{figure}

Interestingly, the key difference between the two methods (discrete SAOS and OWCh) lies in the speed of computation of the mechanical spectrum, as illustrated in Fig.~\ref{Fig.4}. The advantage of using OWCh comes from the fact that in the SAOS one needs to average over several cycles (we used 3 in the specific case) to obtain reasonable statistics. The averaging is mostly needed at low frequency, with the low frequency calculations always constituting the more time-consuming part of the spectrum calculation. The number of cycles needed for the averaging in SAOS can be reduced by increasing the system size, that is the number of particles $N$, therefore increasing in turn the number of iterations required to compute inter-particle interactions in a Molecular Dynamics code, which is the true limiting factor in the simulations of particle based models. In Fig.~\ref{Fig.4}, we also show that the OWCh performance can be further improved by using multi-core processors, as one would expect. On a 20-core processor we obtained a factor 16. The same gain for the same nodes configuration is obtained also with SAOS, but the gain due to OWCh (by not requiring the averaging over several cycles) will remain. That is, for our calculation the OWCh protocol is 3 times faster than a traditional SAOS discrete frequency sweep on a computer equipped with one single core CPU and up to 50 times faster with a 20 core processor. This demonstrates the power of the OWCh protocol for characterizing computationally the viscoelasticity of model soft gels.

\begin{figure}[t]
\includegraphics[width=\linewidth]{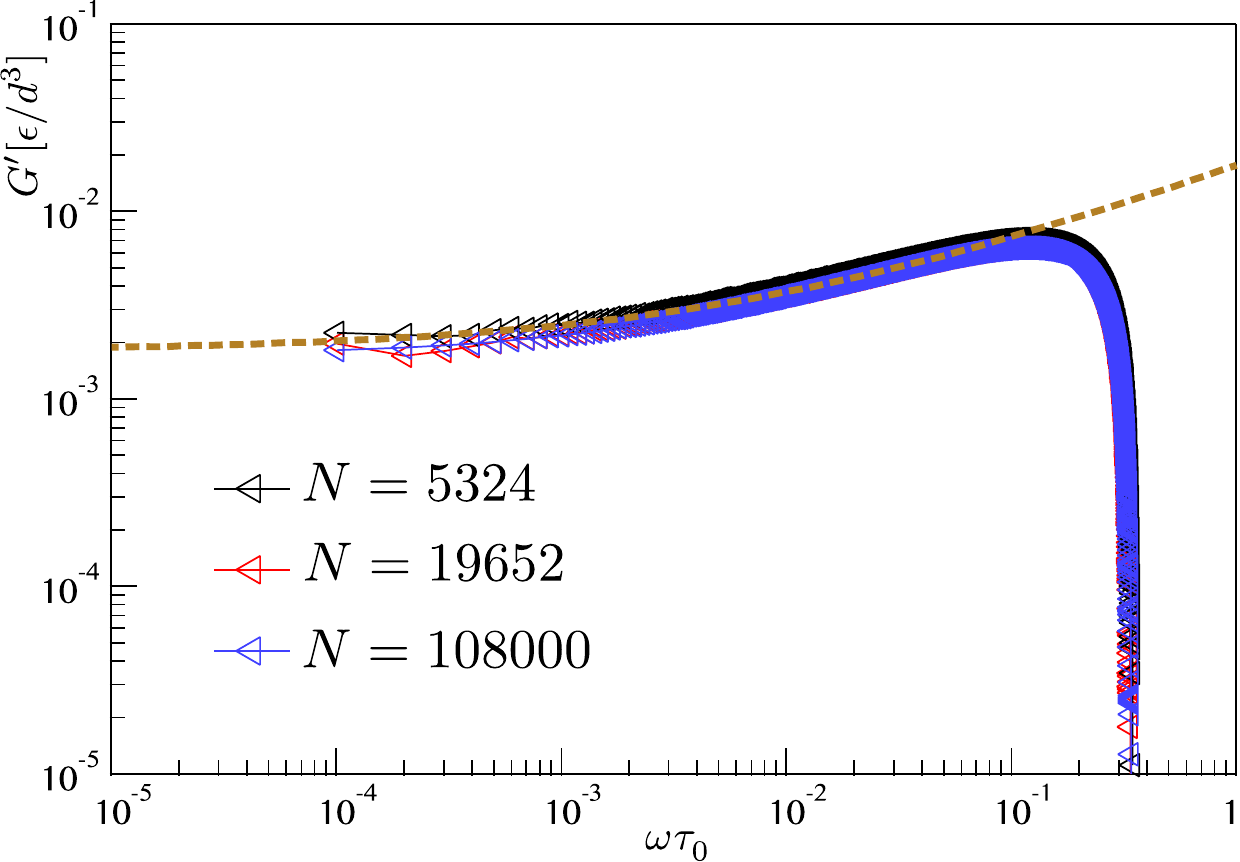}
\caption{Frequency dependence of the storage modulus $G'(\omega)$ for different system size: $N=5324$, 19652 and 108000 particles. Other parameters are set to $\phi\sim8.2\%$, $m=1$ and $\eta_f=0.35$, the dashed brown line represents the best fit of the data by the fractional model introduced in Fig.~\ref{Fig.1}(b) and eq.~(\ref{FDE}), in the intertialess limit of $M\to0$. 
\label{Fig.8}}
\end{figure}

 Possible finite system size effects in the results of the numerical simulations have been ruled out as shown in Fig.~\ref{Fig.8}, which confirms that the OWCh protocol allows to easily span large system sizes over the whole frequency spectrum ($N$ varies between $5 \cdot 10^{3}$ and $10^{5}$). The data also show that the power-law increase of $G'(\omega)$ with frequency and the location of the resonance do not depend on the total number of particles $N$.

\subsection{Fractional Kelvin-Voigt model}
\label{KVM}
 A central feature of the frequency response reported in Fig.~\ref{Fig.3} is the low frequency plateau modulus $G_0$ that resembles the predictions of a Kelvin-Voigt Model (KVM)\cite{Larson:1999}. However, the power-law behavior observed upon increasing the frequency is not captured by a classical Kelvin-Voigt model. The weak power-law-like behavior of soft materials such as food gels and particulate gels can be better captured, in a phenomenological sense, by a spring-pot element, which interpolates between a spring and a dashpot. Such a spring-pot element, which was originally introduced by Scott Blair\cite{Blair1944,Blair1944a} and has recently been applied with success to a broad variety of soft viscoelastic materials\cite{Jaishankar:2013,Wagner:2017} can be represented in terms of a fractional derivative that relates the stress $\sigma$ and the strain $\gamma$ as follows:
\begin{equation}\label{BLAIR}
\sigma = \mathds{V}\frac{d^{\alpha}\gamma}{dt^{\alpha}}
\end{equation}
where  $\alpha$ is a dimensionless exponent $(0\leq\alpha\leq 1)$ and $\mathds{V}$ is referred to as a quasi-property with dimension of Pa.s$^{\alpha}$. Here the operator $d^\alpha/dt^\alpha$ is the Caputo fractional derivative defined as\cite{Podlubny1998}:
\begin{equation}\label{caputo}
\frac{d^\alpha}{dt^\alpha}f(t)=\frac{1}{\Gamma(1-\alpha)}\int_{0}^{t}(t-t')^{-\alpha}f(t')dt'
\end{equation}
with $\Gamma$ the Euler gamma function. This mechanical model captures, in a relatively simple way, a continuous relaxation spectrum that results in a relaxation modulus that decays as a power law in time $G(t)= \mathds{V}t^{-\alpha}/\Gamma(1-\alpha)$.\\ 
To describe the complex modulus of our weak particulate gel, we use therefore a fractional Kelvin-Voigt Model (FKVM) built upon a spring-pot element in parallel with an elastic spring. A functionally-identical modified Kelvin-Voigt model was adopted by Curro and Pincus to describe power-law relaxation in incompletely crosslinked elastomeric systems\cite{Curro1983}. To be able to reproduce the full spectrum, including the resonance that arises from the particle inertia, we have constructed a FKVM in which the spring-pot and the elastic spring are connected in series to an inertial element $M$ that has the dimensions of a mass per unit length [Fig.~\ref{Fig.1}(b)]. This mechanical system has three lumped parameters that physically represent the gel elasticity (the spring $G_0$), the power-law viscous dissipation in the gel (the springpot $\mathbb{V},\alpha$), and the effective inertia contribution to the spectrum (the mass element $M$).

The equation of motion (per unit length) for this mechanical system is a 2$^{nd}$ order fractional differential equation connecting the strain $\gamma(t)$ to the total stress in the system that reads:
\begin{equation}\label{FDE}
M\frac{d^2\gamma}{dt^2}=-\mathds{V}\frac{d^{\alpha}\gamma}{dt^{\alpha}}-G_0\gamma+ \sigma(t)
\end{equation}
Knowing that the Fourier transform  $\mathscr{F}$ of a fractional derivative is given by\cite{Schiessel1995a}:
\begin{equation}\label{fourier}
\mathscr{F}\left\{\frac{d^\alpha}{dt^\alpha}f(t);\omega\right\}=(i\omega)^\alpha\tilde{f}(\omega)
\end{equation}
one can transform the equation of motion (Eq. \ref{FDE}) to the following form:
\begin{equation}
\tilde{\sigma}(\omega)=(-M\omega^2+\mathds{V}(i\omega)^{\alpha}+G_0)\tilde{\gamma}(\omega)
\end{equation}
where the tilde denotes the Fourier transform. Using the continuous expressions displayed in Eq.~\ref{FFT}, we can derive analytical predictions for the elastic and viscous modulus of the FKVM from the real and imaginary components of the transform:
\begin{subequations}\label{gpgdpp}
\begin{equation}
\frac{G'(\omega)}{G_0}=1-\left(\frac{\omega}{\omega_n}\right)^2 + \left(\frac{\omega}{\omega_n}\right)^{\alpha}\xi \cos \left(\alpha\frac{\pi}{2}\right) \end{equation}
\begin{equation}
\frac{G''(\omega)}{G_0}= \left(\frac{\omega}{\omega_n}\right)^{\alpha}\xi \sin \left(\alpha\frac{\pi}{2}\right)
\end{equation}
\end{subequations}
where $\omega_n\equiv\sqrt{G_0/M}$ is the natural resonance frequency of the mass-spring elements and $\xi\equiv\mathbb{V}/\sqrt{M^\alpha G_0^{2-\alpha}}$ is the dimensionless damping ratio that describes the overall power-law dissipative behavior in the system. These predictions from the fractional Kelvin-Voigt model can be used to fit the frequency spectrum obtained by numerical simulations. Indeed, the two equations~(\ref{gpgdpp}a) and (\ref{gpgdpp}b) with the following set of parameters: $G_0=5.10^{-4} \epsilon/d^3$, $\mathds{V}=$0.025, $M=0.12$, and $\alpha=0.53$ capture very well the entire frequency dependence of the elastic and viscous modulus (see black dashed lines in Fig.~\ref{Fig.3}). We only observe small deviations in the viscous modulus at high frequencies, corresponding to the timescales over which the single particle motion is not completely overdamped (see Eq.\ref{shear}) and depending on the specific shape of the interaction potential. The fractional element of the FKV model ($\mathbb{V},\alpha$) accounts for the power-law behavior observed in both the elastic and viscous moduli (corresponding to the terms scaling as $\omega^\alpha$ in Eq.~(\ref{gpgdpp}), while the elastic element ($G_0$) contributes a constant value to the elastic modulus, which dominates at low frequencies. Finally, the effective inertia $M$ of the $N$ particles in the simulation box introduces a mechanical resonance by contributing the term $-G_0(\omega/\omega_n)^2$ to the elastic modulus. This can lead to an unphysical sign change in the apparent elastic modulus that we discuss further below (see Appendix~\ref{appendix3})  The effective inertia $M$ has units of mass per unit length and $G_0$ has dimensions of an energy per volume, hence they will be proportional, respectively, to $m_0/d$ and  $\epsilon/d^3$ in the underlying microstructural computational model. Therefore the natural resonance frequency $\omega_n$ is proportional to the inverse of the characteristic time scale $\tau_0=\sqrt{m_0d^2/\epsilon}$, while the proportionality factor in the scaling of $G_0$ with $\epsilon/d^3$ \textit{a priori} results in a non-trivial way from the structural connectivity of the gel network and the particle volume fraction. In the present work we focus only on the scaling of $G_0$ with $\epsilon/d^{3}$ and the more complicated question of the dependence of $G_{0}$ on the structure of the particulate gel will be subject of future work.

The power-law scaling of the viscous modulus with the frequency $G''(\omega)\sim \omega^\alpha$ (Eq.~\ref{gpgdpp}b) can be understood following a rationale developed by Bagley and Torvik\cite{Bagley:1983,Wharmby:2013} using a modified Rouse theory for soft polymeric materials. The relaxation spectrum can be viewed as a sum of Rouse-like relaxation modes, with $G_0\propto nk_B\mathcal{T}$ and $\tau_1\propto \eta/G_0$, where $n$ is polymer concentration, $\tau_1$ is the Rouse time and $\eta$ represents the viscous contribution of the underlying structure. The summed response of the individual relaxation modes leads to a power-law mechanical response that can be represented in terms of a fractional element with a quasi-property  $\mathbb{V}\propto n k_B\mathcal{T}\tau_1^{\alpha} \propto G_0(\eta/G_0)^{\alpha}$, with $\alpha=1/2$. Following the same general argument, we can deduce the following expected scaling for the numerical gel: $\mathbb{V}\propto G_0(\eta_f d/G_0)^{\alpha}=(\eta_f d)^\alpha G_0^{1-\alpha}$, where we have assumed that for a fixed gel topology the viscous contribution of the gel structure to the response is $\eta \propto \eta_{f} d$. We can now directly test these scaling relationships by varying the particle mass $m$ and the viscous damping $\eta_f$ in the numerical simulations of the same gel structure (keeping all other parameters fixed), and subsequently determining the frequency-dependent complex modulus $G^*(\omega)$ of the gel with the OWCh method.

\begin{figure}[!b]
\includegraphics[width=1\linewidth]{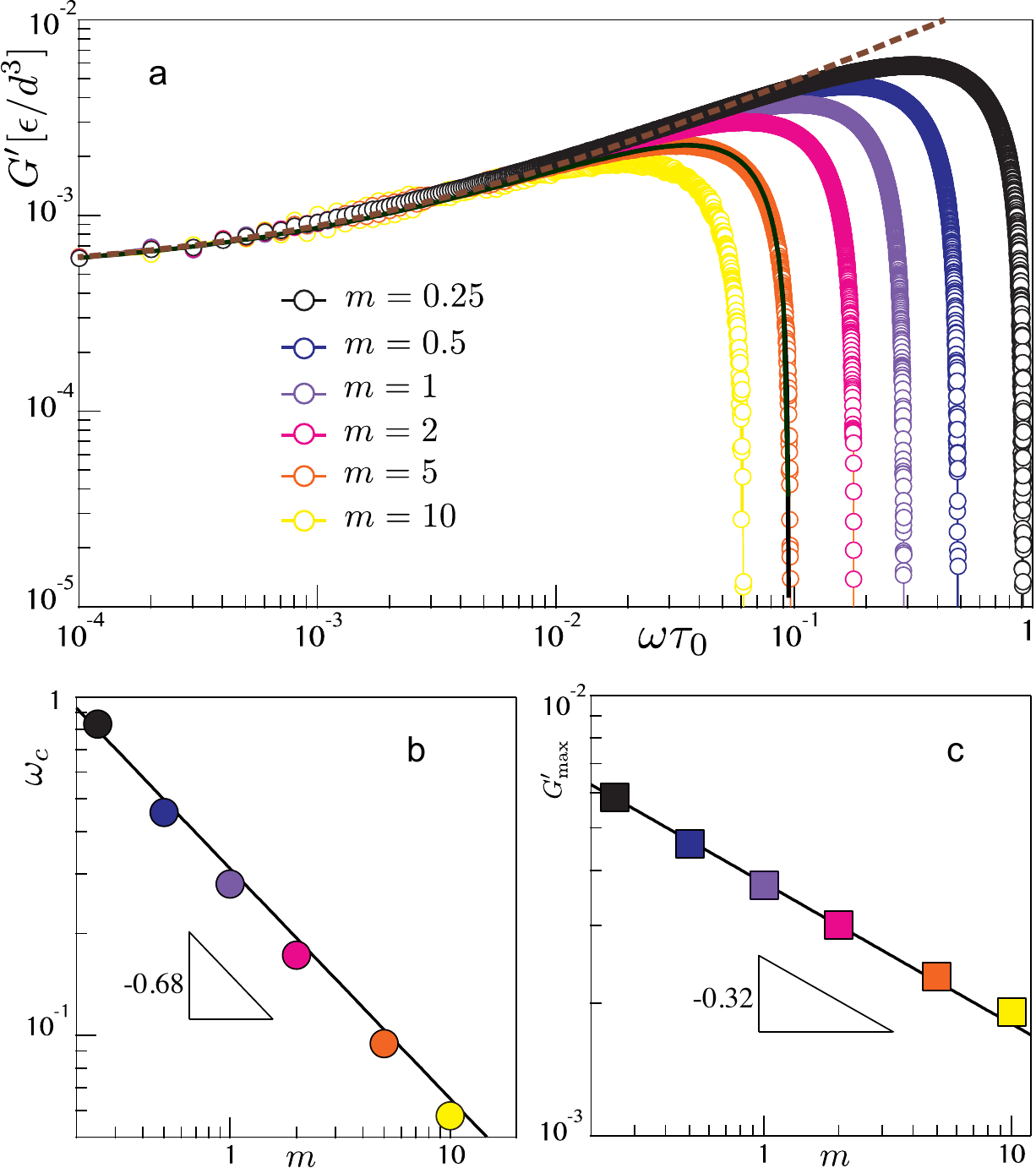}
\caption{(a) Frequency dependence of the storage modulus $G'(\omega)$ for different values of the particle mass $m$ varied over more than a decade with a fixed value of the viscous damping $\eta_f=0.5$. The solid black curve corresponds to equation~(\ref{gpgdpp}a) for the data ($m=5$). Similar qualities of fit are obtained for each data set, and the corresponding values of the model parameters are given in the text. The dashed line represents  to the best fit of the data by the fractional model introduced in Fig. 1(b), in the limit of $M\to0$. (b) Scaling of the critical frequency $\omega_c$ at which the storage modulus passes through a maximum vs the particle mass. (c) Evolution of the maximum value of the storage modulus $G'_{\rm{max}}$ vs the mass $m$. The two black continuous lines correspond to the scaling prediction by the fractional model (see text).
\label{Fig.5}} 
\end{figure} 

\subsection{Frequency Dependency of Gel Viscoelasticity}
\label{dependency}


The viscoelastic properties of gels that have different particle mass, while keeping the viscous damping $\eta_f$ in the equation of motion for each particle constant, are reported in Fig.~\ref{Fig.5}(a). 
Together, the data of Fig.~\ref{Fig.5}(a) and Fig.~\ref{Fig.8} clearly prove that the resonance in the spectrum obtained from the simulations is due to the single particle inertia.The resonance frequency $\omega_c$ changes indeed as we change $m$ (while keeping the total number of particles $N$ constant) and shifts to higher frequency upon decreasing $m$, consistent with the limit $M \rightarrow 0$ obtained for the FKVM model and shown in Fig.~\ref{Fig.8}. 
$\omega_c$ does not change, instead, if we change the total mass of the system $\bar{M} = Nm$ by changing only $N$ while keeping $m$ fixed, whereas it does change even when we change $m$ and keep $\bar{M}$ constant (see also Fig.~\ref{Fig.10} in the Appendix \ref{appendix1}).


Both the maximum $G'_{\rm{max}}$ and $\omega_c$ decrease as power-laws for increasing particle mass $m$ with the  following scalings $G'_{\rm{max}} \sim m^{-2/3}$ and $\omega_c \sim m^{-1/3}$  [Fig.~\ref{Fig.5}(b) and (c) where the scaling exponents where obtained by a self-consistent fitting procedure]. 

\begin{figure}[!th]
\includegraphics[width=1\linewidth]{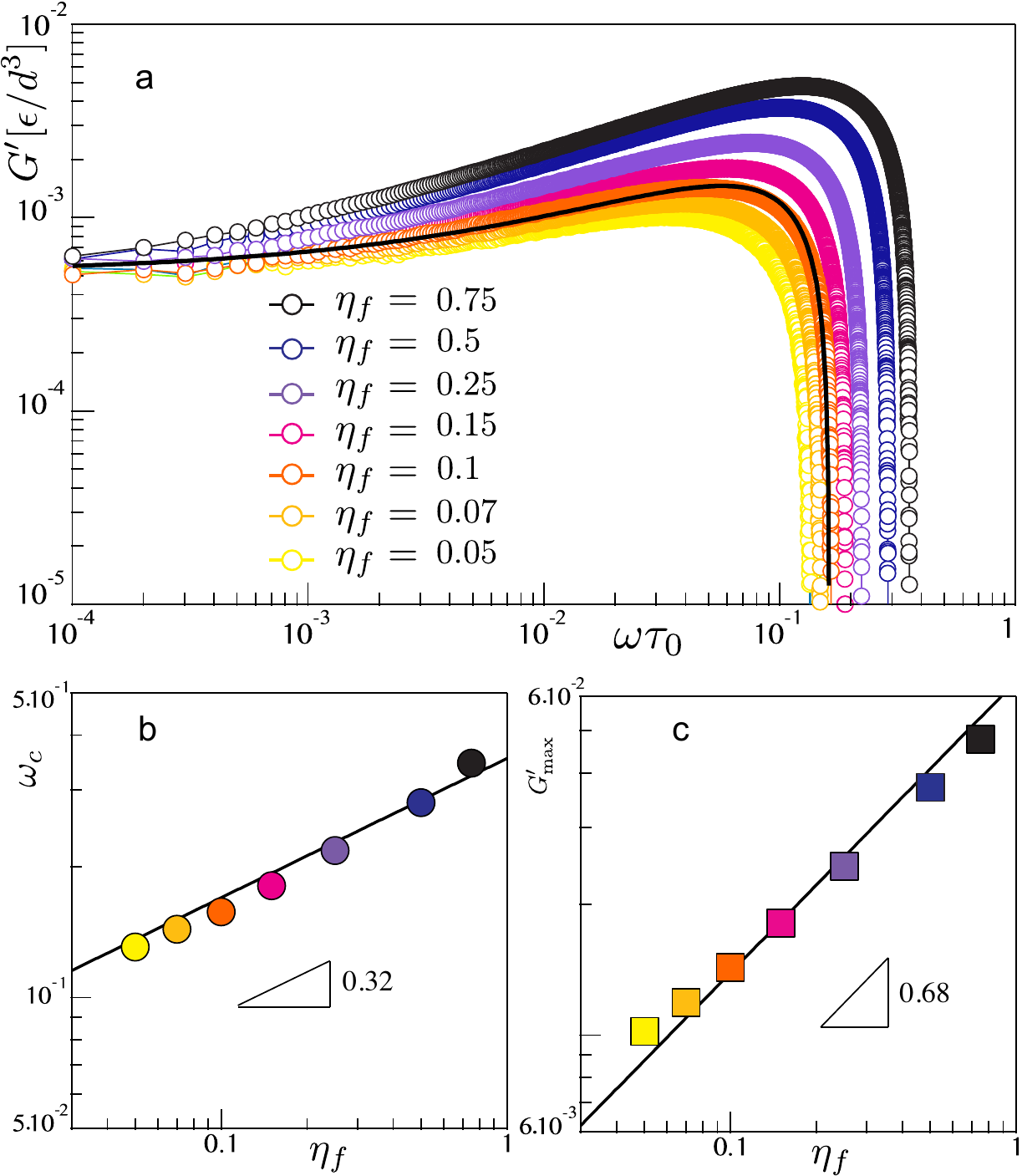}
\caption{(a) Frequency dependence of the storage modulus $G'(\omega)$ for different values of the viscous damping $\eta_f$ with a fixed value of the particle mass, $m=1$. The solid black curve corresponds to equation~(\ref{gpgdpp}a) for the data ($\eta_f=0.1$). Similar goodness of fits are obtained for each data set, and the corresponding values of the model parameters are given in the text. (b) Scaling of the critical frequency $\omega_c$ at which the storage modulus is maximum as the viscous damping  $\eta_f$ is varied. (c) Scaling of the maximum value of the storage modulus $G'_{\rm{max}}$ vs the viscous damping $\eta_f$. The two black continuous lines correspond to the scaling prediction from the fractional model (obtained by a self-consistent fitting procedure).
\label{Fig.6}} 
\end{figure}

Additional simulations show that the elastic modulus of gels prepared at constant particle mass but with different values of the viscous damping $\eta_f$ also show a similar shape with a maximum at a finite frequency [Fig.~\ref{Fig.6}(a)]. However, both the maximum value of the elastic modulus $G'_{\rm{max}}$ and the corresponding critical frequency $\omega_c$ increase for increasing viscous damping $\eta_f$, and follow power-law scalings of the form $G'_{\rm{max}} \sim \eta_f^{2/3}$ and $\omega_c\sim\eta_f^{1/3}$ over the following range of viscous damping $0.05\sqrt{\epsilon m_0/d^2}\leq\eta_f\leq0.75\sqrt{\epsilon m_0/d^2}$.

From a physical viewpoint we can see that by increasing the particle mass, the critical frequency above which inertia effects dominate the gel's response, shifts systematically to smaller values and concomitantly the maximum in the gel elastic modulus decreases. Conversely, increasing viscous damping leads to the opposite trend and delays the onset of significant inertial effects to higher frequencies, which is consistent with the larger values of $G'_{\rm{max}}$ for more heavily damped systems shown in Fig \ref{Fig.6}. 
 
Pursuing the comparison between the numerical gel composed of $N=19652$ particles and the 3 parameter fractional Kelvin-Voigt model with inertia, we can find analytical predictions from Eq.~(\ref{gpgdpp}) for both the maximum elastic modulus $G'_{\rm{max}}$ and the corresponding frequency $\omega_c$. The maximum in $G'(w)$ is defined by $dG'(\omega)/d\omega|_{\omega_c}=0$. Using Eq.~(\ref{gpgdpp}a) we find the following expression for the critical frequency:
\begin{equation}\label{omegac}
\omega_c=\omega_n \left(\frac{\alpha \xi\cos(\alpha\pi/2)}{2}\right)^{1/(2-\alpha)}
\end{equation}
which combined with Eq.~(\ref{gpgdpp}b), leads to the following expression for the maximum value of the elastic modulus:
\begin{equation}\label{gpm}
\frac{G'_{\rm{max}}}{G_0}=\frac{G'(\omega_c)}{G_0}=1+\left(\frac{2}{\alpha}-1\right)\left(\frac{\alpha \xi\cos(\alpha\pi/2)}{2}\right)^{2/(2-\alpha)}	
\end{equation}
Recalling that $\xi\equiv\mathbb{V}/\sqrt{M^\alpha G_0^{2-\alpha}}$ and using the scaling we proposed in section~\ref{KVM} for the quasi-property, i.e. $\mathbb{V}\propto\eta_f^\alpha G_0^{1-\alpha}$, we can now convert the last two expressions into the following appropriate scaling laws:       
\begin{equation}\label{gpm}
\begin{split}
&\omega_c\sim\sqrt{\frac{G_0}{M}}\left(\frac{\alpha\cos(\alpha\pi/2)\eta_f^\alpha G_0^{1-\alpha}}{2\sqrt{M^\alpha G_0^{2-\alpha}}}\right)^{1/(2-\alpha)}\\
& \frac{G'_m}{G_0}-1\sim\left(\frac{2}{\alpha}-1\right) \left(\frac{\alpha\cos(\alpha\pi/2)\eta_f^\alpha G_0^{1-\alpha}}{2\sqrt{M^\alpha G_0^{2-\alpha}}}\right)^{2/(2-\alpha)}
\end{split}.
\end{equation}
When considering variations in particle mass and viscous damping, these scaling laws can be reduced to $\omega_c\sim \eta_f^{\alpha/(2-\alpha)} M^{-1/(2-\alpha)}$ and $G'_{\rm{max}}\sim \eta_f^{2\alpha/(2-\alpha)} M^{-\alpha/(2-\alpha)}$. Substituting $\alpha=0.5$, as determined by the fit of this model to the linear rheology in Fig.~\ref{Fig.3}, we find scaling laws with the same numerical exponents as the ones measured on the numerical gels and reported in Fig.~\ref{Fig.5}(b) and (c), and Fig.~\ref{Fig.6}(b) and (c).

\begin{figure}[!t]
\includegraphics[width=1\linewidth]{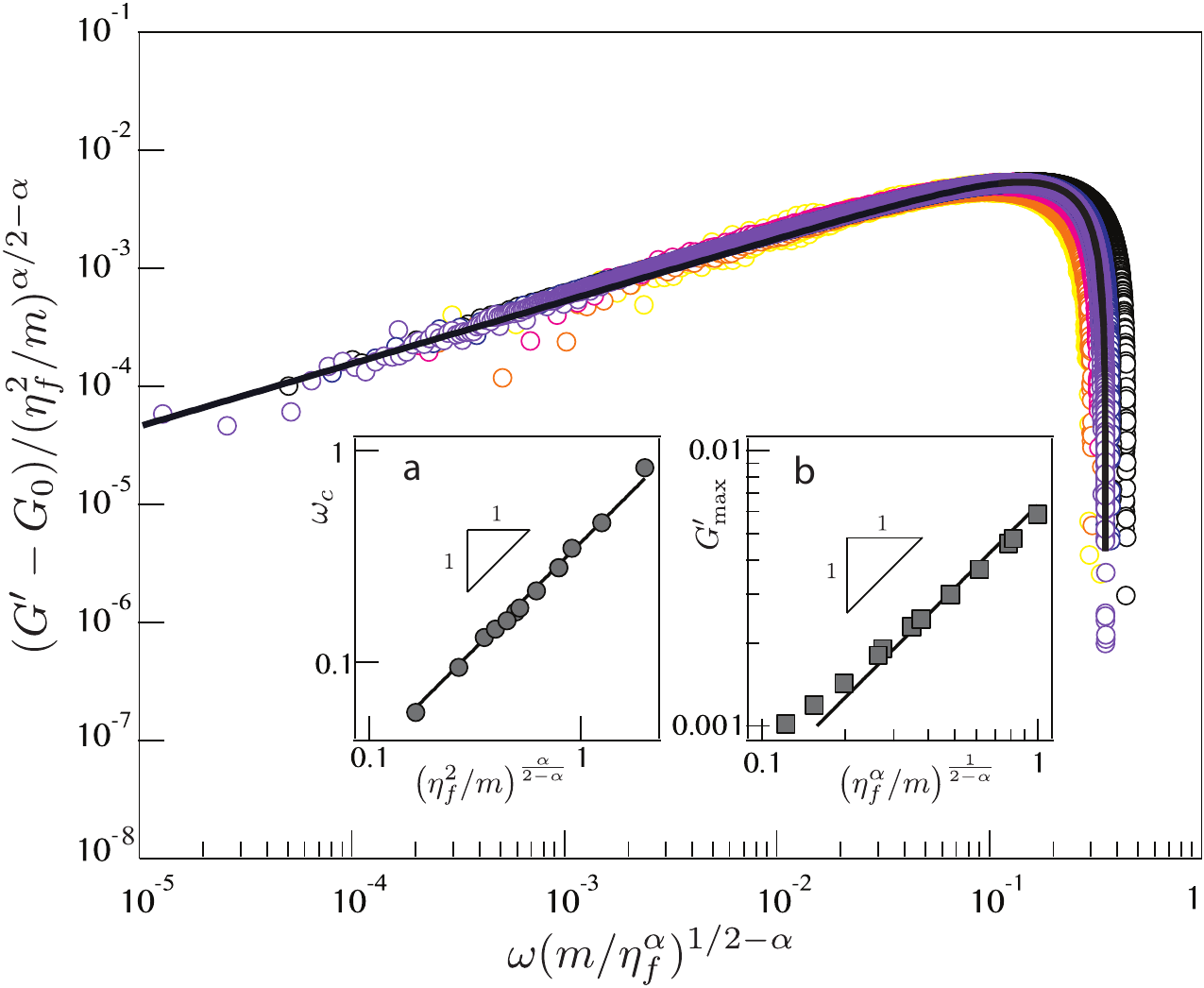}
\caption{Normalized frequency dependence of the storage modulus corresponding to the data presented in Fig.~\ref{Fig.5}(a) and~\ref{Fig.6}(a). Both axes are normalized using the scaling laws predicted with the fractional Kelvin-Voigt model. Insets: (a) Critical frequency $\omega_c$ at which the elastic modulus is determined to be maximum vs the scaling estimate from the fractional model. (b) Same for the maximum of the elastic modulus. 
\label{Fig.7}} 
\end{figure}

We can now use Eq.~\ref{gpm} to rescale the elastic moduli computed for all of the simulated particulate gels onto a single dimensionless plot. In Fig.~\ref{Fig.7} we show the elastic modulus for all of the numerical gels prepared by varying either the particle mass $m$ or the viscous damping $\eta_f$. The horizontal and vertical axis have been normalized by the scaling expressions extracted from the fractional model predictions (Eq.~\ref{gpm}). The collapse of all the numerical data onto a single master curve demonstrates that the agreement with the FKVM lumped parameter model, despite its simplicity, is not just coincidental. The FKV model, in fact, correctly captures the relationships between the different model parameters characterizing the gel (i.e.~the particle mass and the viscous damping coefficient) and correctly predicts how the relaxation spectrum of the numerically-simulated gel depends on these parameters over a wide range of frequencies.


\section{Discussion and conclusion}
\label{discussion}
We have shown that the newly developed OWCh technique for the efficient sampling of the viscoelastic spectrum of complex materials can be successfully used in computational studies of soft particulate gels. In particular, we have demonstrated that the performance advantages of OWCh overcome the long-standing challenges resulting from the length of the numerical tests required and the spectral leakage. On this basis, the OWCh protocol offers potentially broad impact on the fast growing body of computational rheological studies and can tremendously enhance their capability to complement experiments. The advantages brought forward by OWCh have allowed us, in this study, to obtain a quantitative link between the viscoelastic spectrum of a model soft gel and a mesoscopic constitutive model, the fractional Kelvin-Voigt mechanical model. While this class of fractional models has been proposed to correctly capture distinctive features of the viscoelastic spectrum of complex fluids in many different contexts\cite{Friedrich:1999,Ng:2008,Nicolle:2010,Jaishankar:2013}, this is the first time, to our knowledge, that a quantitative connection with a microscopic computational model has been established. Hence our work paves the way to using FKVM to gain new physical insight into the connection between the microscopic physical processes on the particulate level and the resulting macroscopic viscoelastic properties as well as into the complexity of the gel rheological response.

One specific outcome of this computational study shows that varying the inertia of the individual particles and the viscous damping provided by the surrounding solvent changes the position of the resonance frequency $\omega_{c}$ and the corresponding maximum value of the storage modulus $G'_{max}$ exactly as predicted by the FKVM scaling through changes in the inertial element $M$ and the quasi-property $\mathbb{V}$, without changing $G_{0}$ and $\alpha$ (see Fig.~\ref{Fig.1} and Figs.~\ref{Fig.5} - \ref{Fig.7}). 

The inertial element $M$ (with the dimensions of a mass per unit length or linear density) has been introduced in the FKVM to account for the combined effects of particle inertia in the microscopic gel model, since the equations of motion, solved in our molecular dynamics study in the over-damped limit, explicitly contain inertia. In a first instance, $M$ could have been thought of simply as the total mass per unit length in the system (i.e., $\bar{M}/d$ - with $\bar{M}=Nm$). When we consider that the natural resonance of the FKVM is $\omega_{n} = \sqrt{G_{0}/M}$, the fits to the simulation data in Figs.~\ref{Fig.8} and \ref{Fig.5} (see also Fig.\ref{Fig.10} in the Appendix) show that both the resonance position and the low frequency modulus $G_{0}$ do not depend on the number of particles $N$ for the same gel (for which we change only $m$, $\eta_{f}$ or $N$), indicating therefore that $M$ depends on $\bar{M}$ but not on the discretization $N$. Hence one can conclude that $M = f(M/Nd)$, that is, $M$ should be rather thought of as a mass per unit length distributed over the $N$ particles. This is confirmed by the scaling obtained in Fig.\ref{Fig.8} for the numerical model, where $m = \bar{M} /N$ appears with the same dependence (a power law) as $M$. Interestingly, the fact that the resonance $\omega_{c}$ in the numerical model scales with the particle mass $m$ with a power-law (as it does on $M$) can be seen as if the effective mass controlling the resonance in the FKVM model would depend on $\bar{M}/N$ with a power law, and hence as if the effective mass density in the gel leading to the parameter $M$ in the FKVM was distributed in a fractal way among the particles, as in a mass fractal\cite{Shih:1990}. As a consequence, in spite of the fact that the parameter $M$ is included in the model only to reproduce a global feature (the inertia) that is not resolvable for real colloidal gels, it provides interesting clues to the physical meaning of the form of the effective mesoscopic constitutive model.

Our study also allows us to conclude that $G_{0}$ and $\alpha$, on the other hand, must be essentially determined by the only features that we do not vary in this study: the connectivity and the topology of the gel network. Overall our  findings are not inconsistent with the idea that a hyperscaling relationship may exist that links the fractional exponent $\alpha$ to the fractal dimension of the gel network\cite{Bremer:1989,Curtis:2013,Hung:2015}. Nevertheless, we note that the microstructure of the gel considered here (i.e. the organization of the gel particles in space), while certainly porous and heterogeneous, does not obviously display self-similarity over a range of length-scales sufficient to justify such type of connection\cite{Colombo:2014b}. However, in all cases, our results support a connection between the extended power-law regime observed in the viscoelastic spectrum and the disordered gel topology, which features extended or quasi-localized soft modes. Building on the present study, systematic variations in the gel topology and investigation of the modal dynamics over a range of different lengths and timescales can help bring additional understanding of the viscoelastic spectra and of the FKVM parameters.

The viscoelastic response we compute in the molecular dynamics simulations uses the part of the stress that specifically comes from the inter-particle interactions (Eq.~\ref{vstress}) without considering explicitly the flow of the solvent and the hydrodynamic interactions, while we also neglect the role of thermal fluctuations. Hence our results suggest that the complex topology of the particulate gel network alone, disordered and poorly connected, has a mechanical equivalent on much larger length scales that is of the form predicted by a fractional Kelvin-Voigt element, in which a power-law frequency-dependence (characterized by the exponent $\alpha$) arises. Such insight could not be easily gained through directly comparing the same FKVM with experiments, since disentangling different contribution to macroscopic stresses is hard in bulk rheology experiments. These findings show that the extended power-law regimes typically detected in the viscoelastic spectra of soft materials can already emerge from complex stress transmission through the gel structure, even in the absence of a long-range hydrodynamic coupling. The power-law frequency dependence in the dissipation may originate from the damping of extended soft modes, since thermal fluctuations are neglected. These modes are essentially determined by the disordered network topology and involve length-scales larger than the individual particle size. In future studies, we plan to build on the results of this analysis: we will systematically explore the role of the volume fraction of the solid phase in the gel network and use the OWCh protocol to disentangle the roles of thermal fluctuations and of the structural topological constraints in the viscoelastic response of soft gels. 

\section*{References}
\bibliography{biblio-edg}

\begin{thebibliography}{94}%
\makeatletter
\providecommand \@ifxundefined [1]{%
 \@ifx{#1\undefined}
}%
\providecommand \@ifnum [1]{%
 \ifnum #1\expandafter \@firstoftwo
 \else \expandafter \@secondoftwo
 \fi
}%
\providecommand \@ifx [1]{%
 \ifx #1\expandafter \@firstoftwo
 \else \expandafter \@secondoftwo
 \fi
}%
\providecommand \natexlab [1]{#1}%
\providecommand \enquote  [1]{``#1''}%
\providecommand \bibnamefont  [1]{#1}%
\providecommand \bibfnamefont [1]{#1}%
\providecommand \citenamefont [1]{#1}%
\providecommand \href@noop [0]{\@secondoftwo}%
\providecommand \href [0]{\begingroup \@sanitize@url \@href}%
\providecommand \@href[1]{\@@startlink{#1}\@@href}%
\providecommand \@@href[1]{\endgroup#1\@@endlink}%
\providecommand \@sanitize@url [0]{\catcode `\\12\catcode `\$12\catcode
  `\&12\catcode `\#12\catcode `\^12\catcode `\_12\catcode `\%12\relax}%
\providecommand \@@startlink[1]{}%
\providecommand \@@endlink[0]{}%
\providecommand \url  [0]{\begingroup\@sanitize@url \@url }%
\providecommand \@url [1]{\endgroup\@href {#1}{\urlprefix }}%
\providecommand \urlprefix  [0]{URL }%
\providecommand \Eprint [0]{\href }%
\providecommand \doibase [0]{http://dx.doi.org/}%
\providecommand \selectlanguage [0]{\@gobble}%
\providecommand \bibinfo  [0]{\@secondoftwo}%
\providecommand \bibfield  [0]{\@secondoftwo}%
\providecommand \translation [1]{[#1]}%
\providecommand \BibitemOpen [0]{}%
\providecommand \bibitemStop [0]{}%
\providecommand \bibitemNoStop [0]{.\EOS\space}%
\providecommand \EOS [0]{\spacefactor3000\relax}%
\providecommand \BibitemShut  [1]{\csname bibitem#1\endcsname}%
\let\auto@bib@innerbib\@empty
\bibitem [{\citenamefont {Mezzenga}\ \emph {et~al.}(2005)\citenamefont
  {Mezzenga}, \citenamefont {Schurtenberger}, \citenamefont {Burbidge},\ and\
  \citenamefont {Michel}}]{Mezzenga:2005}%
  \BibitemOpen
  \bibfield  {author} {\bibinfo {author} {\bibfnamefont {R.}~\bibnamefont
  {Mezzenga}}, \bibinfo {author} {\bibfnamefont {P.}~\bibnamefont
  {Schurtenberger}}, \bibinfo {author} {\bibfnamefont {A.}~\bibnamefont
  {Burbidge}}, \ and\ \bibinfo {author} {\bibfnamefont {M.}~\bibnamefont
  {Michel}},\ }\bibfield  {title} {\enquote {\bibinfo {title} {Understanding
  foods as soft materials},}\ }\href@noop {} {\bibfield  {journal} {\bibinfo
  {journal} {Nature Materials}\ }\textbf {\bibinfo {volume} {4}},\ \bibinfo
  {pages} {729--740} (\bibinfo {year} {2005})}\BibitemShut {NoStop}%
\bibitem [{\citenamefont {Lu}\ \emph {et~al.}(2008)\citenamefont {Lu},
  \citenamefont {Zaccarelli}, \citenamefont {Ciulla}, \citenamefont
  {Schofield}, \citenamefont {Sciortino},\ and\ \citenamefont
  {Weitz}}]{Lu:2008}%
  \BibitemOpen
  \bibfield  {author} {\bibinfo {author} {\bibfnamefont {P.}~\bibnamefont
  {Lu}}, \bibinfo {author} {\bibfnamefont {E.}~\bibnamefont {Zaccarelli}},
  \bibinfo {author} {\bibfnamefont {F.}~\bibnamefont {Ciulla}}, \bibinfo
  {author} {\bibfnamefont {A.~B.}\ \bibnamefont {Schofield}}, \bibinfo {author}
  {\bibfnamefont {F.}~\bibnamefont {Sciortino}}, \ and\ \bibinfo {author}
  {\bibfnamefont {D.~A.}\ \bibnamefont {Weitz}},\ }\bibfield  {title} {\enquote
  {\bibinfo {title} {Gelation of particles with short-range attraction},}\
  }\href@noop {} {\bibfield  {journal} {\bibinfo  {journal} {Nature}\ }\textbf
  {\bibinfo {volume} {453}},\ \bibinfo {pages} {499--503} (\bibinfo {year}
  {2008})}\BibitemShut {NoStop}%
\bibitem [{\citenamefont {Conrad}\ and\ \citenamefont
  {Lewis}(2008)}]{Conrad:2010}%
  \BibitemOpen
  \bibfield  {author} {\bibinfo {author} {\bibfnamefont {J.~C.}\ \bibnamefont
  {Conrad}}\ and\ \bibinfo {author} {\bibfnamefont {J.~A.}\ \bibnamefont
  {Lewis}},\ }\bibfield  {title} {\enquote {\bibinfo {title} {Structure of
  colloidal gels during microchannel flow},}\ }\href@noop {} {\bibfield
  {journal} {\bibinfo  {journal} {Langmuir}\ }\textbf {\bibinfo {volume}
  {24}},\ \bibinfo {pages} {7628--7634} (\bibinfo {year} {2008})}\BibitemShut
  {NoStop}%
\bibitem [{\citenamefont {Helgeson}\ \emph {et~al.}(2012)\citenamefont
  {Helgeson}, \citenamefont {Moran}, \citenamefont {An},\ and\ \citenamefont
  {Doyle}}]{Helgeson:2012}%
  \BibitemOpen
  \bibfield  {author} {\bibinfo {author} {\bibfnamefont {M.}~\bibnamefont
  {Helgeson}}, \bibinfo {author} {\bibfnamefont {S.}~\bibnamefont {Moran}},
  \bibinfo {author} {\bibfnamefont {H.}~\bibnamefont {An}}, \ and\ \bibinfo
  {author} {\bibfnamefont {P.}~\bibnamefont {Doyle}},\ }\bibfield  {title}
  {\enquote {\bibinfo {title} {Mesoporous organohydrogels from thermogelling
  photocrosslinkable nanoemulsions},}\ }\href@noop {} {\bibfield  {journal}
  {\bibinfo  {journal} {Nature Materials}\ }\textbf {\bibinfo {volume} {11}},\
  \bibinfo {pages} {344--352} (\bibinfo {year} {2012})}\BibitemShut {NoStop}%
\bibitem [{\citenamefont {Gibaud}\ \emph {et~al.}(2013)\citenamefont {Gibaud},
  \citenamefont {Zaccone}, \citenamefont {{Del Gado}}, \citenamefont {Trappe},\
  and\ \citenamefont {Schurtenberger}}]{Gibaud:2013aa}%
  \BibitemOpen
  \bibfield  {author} {\bibinfo {author} {\bibfnamefont {T.}~\bibnamefont
  {Gibaud}}, \bibinfo {author} {\bibfnamefont {A.}~\bibnamefont {Zaccone}},
  \bibinfo {author} {\bibfnamefont {E.}~\bibnamefont {{Del Gado}}}, \bibinfo
  {author} {\bibfnamefont {V.}~\bibnamefont {Trappe}}, \ and\ \bibinfo {author}
  {\bibfnamefont {P.}~\bibnamefont {Schurtenberger}},\ }\bibfield  {title}
  {\enquote {\bibinfo {title} {Unexpected decoupling of stretching and bending
  modes in protein gels},}\ }\href {\doibase 10.1103/PhysRevLett.110.058303}
  {\bibfield  {journal} {\bibinfo  {journal} {Phys. Rev. Lett.}\ }\textbf
  {\bibinfo {volume} {110}},\ \bibinfo {pages} {058303} (\bibinfo {year}
  {2013})}\BibitemShut {NoStop}%
\bibitem [{\citenamefont {Zhao}(2014)}]{Zhao:2014}%
  \BibitemOpen
  \bibfield  {author} {\bibinfo {author} {\bibfnamefont {X.}~\bibnamefont
  {Zhao}},\ }\bibfield  {title} {\enquote {\bibinfo {title} {Multi-scale
  multi-mechanism design of tough hydrogels: building dissipation into strechy
  networks},}\ }\href@noop {} {\bibfield  {journal} {\bibinfo  {journal} {Soft
  Matter}\ }\textbf {\bibinfo {volume} {10}},\ \bibinfo {pages} {672--687}
  (\bibinfo {year} {2014})}\BibitemShut {NoStop}%
\bibitem [{\citenamefont {Grindy}\ \emph {et~al.}(2015)\citenamefont {Grindy},
  \citenamefont {Learsch}, \citenamefont {Mozhdehi}, \citenamefont {Cheng},
  \citenamefont {Barrett}, \citenamefont {Guan}, \citenamefont {Messersmith},\
  and\ \citenamefont {Holten-Andersen}}]{Grindy:2015}%
  \BibitemOpen
  \bibfield  {author} {\bibinfo {author} {\bibfnamefont {S.}~\bibnamefont
  {Grindy}}, \bibinfo {author} {\bibfnamefont {R.}~\bibnamefont {Learsch}},
  \bibinfo {author} {\bibfnamefont {D.}~\bibnamefont {Mozhdehi}}, \bibinfo
  {author} {\bibfnamefont {J.}~\bibnamefont {Cheng}}, \bibinfo {author}
  {\bibfnamefont {D.}~\bibnamefont {Barrett}}, \bibinfo {author} {\bibfnamefont
  {Z.}~\bibnamefont {Guan}}, \bibinfo {author} {\bibfnamefont {P.}~\bibnamefont
  {Messersmith}}, \ and\ \bibinfo {author} {\bibfnamefont {N.}~\bibnamefont
  {Holten-Andersen}},\ }\bibfield  {title} {\enquote {\bibinfo {title} {Control
  of hierarchical polymer mechanics with bioinspired metal-coordination
  dynamics},}\ }\href@noop {} {\bibfield  {journal} {\bibinfo  {journal}
  {Nature Materials}\ }\textbf {\bibinfo {volume} {14}},\ \bibinfo {pages}
  {1210--1217} (\bibinfo {year} {2015})}\BibitemShut {NoStop}%
\bibitem [{\citenamefont {Ewoldt}, \citenamefont {Hosoi},\ and\ \citenamefont
  {McKinley}(2008)}]{Ewoldt:2008}%
  \BibitemOpen
  \bibfield  {author} {\bibinfo {author} {\bibfnamefont {R.~H.}\ \bibnamefont
  {Ewoldt}}, \bibinfo {author} {\bibfnamefont {A.~E.}\ \bibnamefont {Hosoi}}, \
  and\ \bibinfo {author} {\bibfnamefont {G.~H.}\ \bibnamefont {McKinley}},\
  }\bibfield  {title} {\enquote {\bibinfo {title} {New measures for
  characterizing nonlinear viscoelasticity in large amplitude oscillatory
  shear},}\ }\href@noop {} {\bibfield  {journal} {\bibinfo  {journal} {J.
  Rheol.}\ }\textbf {\bibinfo {volume} {52}},\ \bibinfo {pages} {1427--1458}
  (\bibinfo {year} {2008})}\BibitemShut {NoStop}%
\bibitem [{\citenamefont {Ewoldt}\ \emph {et~al.}(2010)\citenamefont {Ewoldt},
  \citenamefont {Winter}, \citenamefont {Maxey},\ and\ \citenamefont
  {McKinley}}]{Ewoldt:2010}%
  \BibitemOpen
  \bibfield  {author} {\bibinfo {author} {\bibfnamefont {R.~H.}\ \bibnamefont
  {Ewoldt}}, \bibinfo {author} {\bibfnamefont {P.}~\bibnamefont {Winter}},
  \bibinfo {author} {\bibfnamefont {J.}~\bibnamefont {Maxey}}, \ and\ \bibinfo
  {author} {\bibfnamefont {G.~H.}\ \bibnamefont {McKinley}},\ }\bibfield
  {title} {\enquote {\bibinfo {title} {Large amplitude oscillatory shear of
  pseudoplastic and elastoviscoplastic materials},}\ }\href@noop {} {\bibfield
  {journal} {\bibinfo  {journal} {Rheol. Acta}\ }\textbf {\bibinfo {volume}
  {49}},\ \bibinfo {pages} {191--212} (\bibinfo {year} {2010})}\BibitemShut
  {NoStop}%
\bibitem [{\citenamefont {Laurati}, \citenamefont {Egelhaaf},\ and\
  \citenamefont {Petekidis}(2011)}]{Laurati:2011}%
  \BibitemOpen
  \bibfield  {author} {\bibinfo {author} {\bibfnamefont {M.}~\bibnamefont
  {Laurati}}, \bibinfo {author} {\bibfnamefont {S.}~\bibnamefont {Egelhaaf}}, \
  and\ \bibinfo {author} {\bibfnamefont {G.}~\bibnamefont {Petekidis}},\
  }\bibfield  {title} {\enquote {\bibinfo {title} {Nonlinear rheology of
  colloidal gels with intermediate volume fraction},}\ }\href@noop {}
  {\bibfield  {journal} {\bibinfo  {journal} {J. Rheol.}\ }\textbf {\bibinfo
  {volume} {55}},\ \bibinfo {pages} {673--706} (\bibinfo {year}
  {2011})}\BibitemShut {NoStop}%
\bibitem [{\citenamefont {Mao}, \citenamefont {Divoux},\ and\ \citenamefont
  {Snabre}(2016)}]{Mao:2016a}%
  \BibitemOpen
  \bibfield  {author} {\bibinfo {author} {\bibfnamefont {B.}~\bibnamefont
  {Mao}}, \bibinfo {author} {\bibfnamefont {T.}~\bibnamefont {Divoux}}, \ and\
  \bibinfo {author} {\bibfnamefont {P.}~\bibnamefont {Snabre}},\ }\bibfield
  {title} {\enquote {\bibinfo {title} {Normal force controlled rheology applied
  to agar gelation},}\ }\href@noop {} {\bibfield  {journal} {\bibinfo
  {journal} {J. Rheol.}\ }\textbf {\bibinfo {volume} {60}},\ \bibinfo {pages}
  {473--489} (\bibinfo {year} {2016})}\BibitemShut {NoStop}%
\bibitem [{\citenamefont {Jaishankar}\ and\ \citenamefont
  {McKinley}(2013)}]{Jaishankar:2013}%
  \BibitemOpen
  \bibfield  {author} {\bibinfo {author} {\bibfnamefont {A.}~\bibnamefont
  {Jaishankar}}\ and\ \bibinfo {author} {\bibfnamefont {G.~H.}\ \bibnamefont
  {McKinley}},\ }\bibfield  {title} {\enquote {\bibinfo {title} {{Power-law
  rheology in the bulk and at the interface: quasi-properties and fractional
  constitutive equations}},}\ }\href@noop {} {\bibfield  {journal} {\bibinfo
  {journal} {Proc. R. Soc. A}\ }\textbf {\bibinfo {volume} {469}},\ \bibinfo
  {pages} {20120284} (\bibinfo {year} {2013})}\BibitemShut {NoStop}%
\bibitem [{\citenamefont {Helal}, \citenamefont {Divoux},\ and\ \citenamefont
  {McKinley}(2016)}]{Helal:2016}%
  \BibitemOpen
  \bibfield  {author} {\bibinfo {author} {\bibfnamefont {A.}~\bibnamefont
  {Helal}}, \bibinfo {author} {\bibfnamefont {T.}~\bibnamefont {Divoux}}, \
  and\ \bibinfo {author} {\bibfnamefont {G.~H.}\ \bibnamefont {McKinley}},\
  }\bibfield  {title} {\enquote {\bibinfo {title} {Simultaneous rheoelectric
  measurements of strongly conductive complex fluids},}\ }\href@noop {}
  {\bibfield  {journal} {\bibinfo  {journal} {Physical Review Applied}\
  }\textbf {\bibinfo {volume} {6}},\ \bibinfo {pages} {064004} (\bibinfo {year}
  {2016})}\BibitemShut {NoStop}%
\bibitem [{\citenamefont {Aime}\ \emph {et~al.}(2016)\citenamefont {Aime},
  \citenamefont {Ramos}, \citenamefont {Fromental}, \citenamefont {Pr\'evot},
  \citenamefont {Jelinek},\ and\ \citenamefont {Cipelletti}}]{Aime:2016}%
  \BibitemOpen
  \bibfield  {author} {\bibinfo {author} {\bibfnamefont {S.}~\bibnamefont
  {Aime}}, \bibinfo {author} {\bibfnamefont {L.}~\bibnamefont {Ramos}},
  \bibinfo {author} {\bibfnamefont {J.~M.}\ \bibnamefont {Fromental}}, \bibinfo
  {author} {\bibfnamefont {G.}~\bibnamefont {Pr\'evot}}, \bibinfo {author}
  {\bibfnamefont {R.}~\bibnamefont {Jelinek}}, \ and\ \bibinfo {author}
  {\bibfnamefont {L.}~\bibnamefont {Cipelletti}},\ }\bibfield  {title}
  {\enquote {\bibinfo {title} {A stress-controlled shear cell for small-angle
  light scattering and microscopy},}\ }\href@noop {} {\bibfield  {journal}
  {\bibinfo  {journal} {Review of Scientific Instruments}\ }\textbf {\bibinfo
  {volume} {87}},\ \bibinfo {pages} {123907} (\bibinfo {year}
  {2016})}\BibitemShut {NoStop}%
\bibitem [{\citenamefont {Laurati}\ \emph {et~al.}(2017)\citenamefont
  {Laurati}, \citenamefont {Masshoff}, \citenamefont {Mutch}, \citenamefont
  {Egelhaaf},\ and\ \citenamefont {Zaccone}}]{Laurati:2017}%
  \BibitemOpen
  \bibfield  {author} {\bibinfo {author} {\bibfnamefont {M.}~\bibnamefont
  {Laurati}}, \bibinfo {author} {\bibfnamefont {P.}~\bibnamefont {Masshoff}},
  \bibinfo {author} {\bibfnamefont {K.~J.}\ \bibnamefont {Mutch}}, \bibinfo
  {author} {\bibfnamefont {S.~U.}\ \bibnamefont {Egelhaaf}}, \ and\ \bibinfo
  {author} {\bibfnamefont {A.}~\bibnamefont {Zaccone}},\ }\bibfield  {title}
  {\enquote {\bibinfo {title} {Long-lived neighbors determine the rheological
  response of glasses},}\ }\href@noop {} {\bibfield  {journal} {\bibinfo
  {journal} {Physical Review Letters}\ }\textbf {\bibinfo {volume} {118}},\
  \bibinfo {pages} {018002} (\bibinfo {year} {2017})}\BibitemShut {NoStop}%
\bibitem [{\citenamefont {Cipelletti}\ and\ \citenamefont
  {Ramos}(2005)}]{Cipelletti:2005}%
  \BibitemOpen
  \bibfield  {author} {\bibinfo {author} {\bibfnamefont {L.}~\bibnamefont
  {Cipelletti}}\ and\ \bibinfo {author} {\bibfnamefont {L.}~\bibnamefont
  {Ramos}},\ }\bibfield  {title} {\enquote {\bibinfo {title} {Slow dynamics in
  glassy soft matter},}\ }\href@noop {} {\bibfield  {journal} {\bibinfo
  {journal} {J. Phys.: Condens. Matter}\ }\textbf {\bibinfo {volume} {17}},\
  \bibinfo {pages} {R253--R285} (\bibinfo {year} {2005})}\BibitemShut {NoStop}%
\bibitem [{\citenamefont {Mohraz}\ and\ \citenamefont
  {Solomon}(2005)}]{Mohraz:2005}%
  \BibitemOpen
  \bibfield  {author} {\bibinfo {author} {\bibfnamefont {A.}~\bibnamefont
  {Mohraz}}\ and\ \bibinfo {author} {\bibfnamefont {M.}~\bibnamefont
  {Solomon}},\ }\bibfield  {title} {\enquote {\bibinfo {title} {Orientation and
  rupture of fractal colloidal gels during start-up of steady shear flow},}\
  }\href@noop {} {\bibfield  {journal} {\bibinfo  {journal} {J. Rheol.}\
  }\textbf {\bibinfo {volume} {49}},\ \bibinfo {pages} {657--681} (\bibinfo
  {year} {2005})}\BibitemShut {NoStop}%
\bibitem [{\citenamefont {Dibble}, \citenamefont {Kogan},\ and\ \citenamefont
  {Solomon}(2008)}]{Dibble:2008}%
  \BibitemOpen
  \bibfield  {author} {\bibinfo {author} {\bibfnamefont {C.~J.}\ \bibnamefont
  {Dibble}}, \bibinfo {author} {\bibfnamefont {M.}~\bibnamefont {Kogan}}, \
  and\ \bibinfo {author} {\bibfnamefont {M.~J.}\ \bibnamefont {Solomon}},\
  }\bibfield  {title} {\enquote {\bibinfo {title} {Structural origins of
  dynamical heterogeneity in colloidal gels},}\ }\href@noop {} {\bibfield
  {journal} {\bibinfo  {journal} {Physical Review E}\ }\textbf {\bibinfo
  {volume} {77}},\ \bibinfo {pages} {050401} (\bibinfo {year}
  {2008})}\BibitemShut {NoStop}%
\bibitem [{\citenamefont {Divoux}\ \emph {et~al.}(2010)\citenamefont {Divoux},
  \citenamefont {Tamarii}, \citenamefont {Barentin},\ and\ \citenamefont
  {Manneville}}]{Divoux:2010}%
  \BibitemOpen
  \bibfield  {author} {\bibinfo {author} {\bibfnamefont {T.}~\bibnamefont
  {Divoux}}, \bibinfo {author} {\bibfnamefont {D.}~\bibnamefont {Tamarii}},
  \bibinfo {author} {\bibfnamefont {C.}~\bibnamefont {Barentin}}, \ and\
  \bibinfo {author} {\bibfnamefont {S.}~\bibnamefont {Manneville}},\ }\bibfield
   {title} {\enquote {\bibinfo {title} {Transient shear banding in a simple
  yield stress fluid},}\ }\href@noop {} {\bibfield  {journal} {\bibinfo
  {journal} {Phys. Rev. Lett.}\ }\textbf {\bibinfo {volume} {104}},\ \bibinfo
  {pages} {208301} (\bibinfo {year} {2010})}\BibitemShut {NoStop}%
\bibitem [{\citenamefont {Divoux}, \citenamefont {Barentin},\ and\
  \citenamefont {Manneville}(2011)}]{Divoux:2011}%
  \BibitemOpen
  \bibfield  {author} {\bibinfo {author} {\bibfnamefont {T.}~\bibnamefont
  {Divoux}}, \bibinfo {author} {\bibfnamefont {C.}~\bibnamefont {Barentin}}, \
  and\ \bibinfo {author} {\bibfnamefont {S.}~\bibnamefont {Manneville}},\
  }\bibfield  {title} {\enquote {\bibinfo {title} {Stress overshoot in a simple
  yield stress fluid: an extensive study combining rheology and velocimetry},}\
  }\href@noop {} {\bibfield  {journal} {\bibinfo  {journal} {Soft Matter}\
  }\textbf {\bibinfo {volume} {7}},\ \bibinfo {pages} {9335--9349} (\bibinfo
  {year} {2011})}\BibitemShut {NoStop}%
\bibitem [{\citenamefont {Callaghan}(2008)}]{Callaghan:2008}%
  \BibitemOpen
  \bibfield  {author} {\bibinfo {author} {\bibfnamefont {P.~T.}\ \bibnamefont
  {Callaghan}},\ }\bibfield  {title} {\enquote {\bibinfo {title} {Rheo {NMR}
  and shear banding},}\ }\href@noop {} {\bibfield  {journal} {\bibinfo
  {journal} {Rheol. Acta}\ }\textbf {\bibinfo {volume} {47}},\ \bibinfo {pages}
  {243--255} (\bibinfo {year} {2008})}\BibitemShut {NoStop}%
\bibitem [{\citenamefont {Manneville}(2008)}]{Manneville:2008}%
  \BibitemOpen
  \bibfield  {author} {\bibinfo {author} {\bibfnamefont {S.}~\bibnamefont
  {Manneville}},\ }\bibfield  {title} {\enquote {\bibinfo {title} {Recent
  experimental probes of shear banding},}\ }\href@noop {} {\bibfield  {journal}
  {\bibinfo  {journal} {Rheol. Acta}\ }\textbf {\bibinfo {volume} {47}},\
  \bibinfo {pages} {301--318} (\bibinfo {year} {2008})}\BibitemShut {NoStop}%
\bibitem [{\citenamefont {Chan}\ and\ \citenamefont
  {Mohraz}(2013)}]{Chan:2013}%
  \BibitemOpen
  \bibfield  {author} {\bibinfo {author} {\bibfnamefont {H.}~\bibnamefont
  {Chan}}\ and\ \bibinfo {author} {\bibfnamefont {A.}~\bibnamefont {Mohraz}},\
  }\bibfield  {title} {\enquote {\bibinfo {title} {A simple shear cell for the
  direct visualization of step-stress deformation in soft materials},}\
  }\href@noop {} {\bibfield  {journal} {\bibinfo  {journal} {Rheol. Acta}\
  }\textbf {\bibinfo {volume} {52}},\ \bibinfo {pages} {383--394} (\bibinfo
  {year} {2013})}\BibitemShut {NoStop}%
\bibitem [{\citenamefont {Guo}\ \emph {et~al.}(2010)\citenamefont {Guo},
  \citenamefont {Ramakrishnan}, \citenamefont {Harden},\ and\ \citenamefont
  {Leheny}}]{Guo:2010}%
  \BibitemOpen
  \bibfield  {author} {\bibinfo {author} {\bibfnamefont {H.}~\bibnamefont
  {Guo}}, \bibinfo {author} {\bibfnamefont {S.}~\bibnamefont {Ramakrishnan}},
  \bibinfo {author} {\bibfnamefont {J.~L.}\ \bibnamefont {Harden}}, \ and\
  \bibinfo {author} {\bibfnamefont {R.~L.}\ \bibnamefont {Leheny}},\ }\bibfield
   {title} {\enquote {\bibinfo {title} {Connecting nanoscale motion and
  rheology of gel-forming colloidal suspensions},}\ }\href@noop {} {\bibfield
  {journal} {\bibinfo  {journal} {Physical Review E}\ }\textbf {\bibinfo
  {volume} {81}},\ \bibinfo {pages} {050401} (\bibinfo {year}
  {2010})}\BibitemShut {NoStop}%
\bibitem [{\citenamefont {Perge}\ \emph {et~al.}(2014)\citenamefont {Perge},
  \citenamefont {Taberlet}, \citenamefont {Gibaud},\ and\ \citenamefont
  {Manneville}}]{Perge:2014b}%
  \BibitemOpen
  \bibfield  {author} {\bibinfo {author} {\bibfnamefont {C.}~\bibnamefont
  {Perge}}, \bibinfo {author} {\bibfnamefont {N.}~\bibnamefont {Taberlet}},
  \bibinfo {author} {\bibfnamefont {T.}~\bibnamefont {Gibaud}}, \ and\ \bibinfo
  {author} {\bibfnamefont {S.}~\bibnamefont {Manneville}},\ }\bibfield  {title}
  {\enquote {\bibinfo {title} {Time dependence in large amplitude oscillatory
  shear: A rheo-ultrasonic study of fatigue dynamics in a colloidal gel},}\
  }\href@noop {} {\bibfield  {journal} {\bibinfo  {journal} {J. Rheol.}\
  }\textbf {\bibinfo {volume} {58}},\ \bibinfo {pages} {1331--1357} (\bibinfo
  {year} {2014})}\BibitemShut {NoStop}%
\bibitem [{\citenamefont {Tamborini}, \citenamefont {Cipelletti},\ and\
  \citenamefont {Ramos}(2014)}]{Tamborini:2014}%
  \BibitemOpen
  \bibfield  {author} {\bibinfo {author} {\bibfnamefont {E.}~\bibnamefont
  {Tamborini}}, \bibinfo {author} {\bibfnamefont {L.}~\bibnamefont
  {Cipelletti}}, \ and\ \bibinfo {author} {\bibfnamefont {L.}~\bibnamefont
  {Ramos}},\ }\bibfield  {title} {\enquote {\bibinfo {title} {Plasticity of a
  colloidal polycrystal under cyclic shear},}\ }\href@noop {} {\bibfield
  {journal} {\bibinfo  {journal} {Physical Review Letters}\ } (\bibinfo {year}
  {2014})}\BibitemShut {NoStop}%
\bibitem [{\citenamefont {Santos}, \citenamefont {Campanella},\ and\
  \citenamefont {Carignano}(2013)}]{Santos:2013}%
  \BibitemOpen
  \bibfield  {author} {\bibinfo {author} {\bibfnamefont {P.}~\bibnamefont
  {Santos}}, \bibinfo {author} {\bibfnamefont {O.}~\bibnamefont {Campanella}},
  \ and\ \bibinfo {author} {\bibfnamefont {M.}~\bibnamefont {Carignano}},\
  }\bibfield  {title} {\enquote {\bibinfo {title} {Effective attractive range
  and viscoelasticity of colloidal gels},}\ }\href@noop {} {\bibfield
  {journal} {\bibinfo  {journal} {Soft Matter}\ }\textbf {\bibinfo {volume}
  {9}},\ \bibinfo {pages} {709--714} (\bibinfo {year} {2013})}\BibitemShut
  {NoStop}%
\bibitem [{\citenamefont {Park}\ and\ \citenamefont {Ahn}(2013)}]{Park:2013}%
  \BibitemOpen
  \bibfield  {author} {\bibinfo {author} {\bibfnamefont {J.}~\bibnamefont
  {Park}}\ and\ \bibinfo {author} {\bibfnamefont {K.}~\bibnamefont {Ahn}},\
  }\bibfield  {title} {\enquote {\bibinfo {title} {Structural evolution of
  colloidal gels at intermediate volume fraction under start-up of shear
  flow},}\ }\href@noop {} {\bibfield  {journal} {\bibinfo  {journal} {Soft
  Matter}\ }\textbf {\bibinfo {volume} {9}},\ \bibinfo {pages} {11650--11662}
  (\bibinfo {year} {2013})}\BibitemShut {NoStop}%
\bibitem [{\citenamefont {Colombo}\ and\ \citenamefont {{Del
  Gado}}(2014)}]{Colombo:2014}%
  \BibitemOpen
  \bibfield  {author} {\bibinfo {author} {\bibfnamefont {J.}~\bibnamefont
  {Colombo}}\ and\ \bibinfo {author} {\bibfnamefont {E.}~\bibnamefont {{Del
  Gado}}},\ }\bibfield  {title} {\enquote {\bibinfo {title} {Stress
  localization, stiffening, and yielding in a model colloidal gel},}\
  }\href@noop {} {\bibfield  {journal} {\bibinfo  {journal} {J. Rheol.}\
  }\textbf {\bibinfo {volume} {58}},\ \bibinfo {pages} {1089--1116} (\bibinfo
  {year} {2014})}\BibitemShut {NoStop}%
\bibitem [{\citenamefont {Varga}\ and\ \citenamefont
  {Swan}(2015)}]{Varga:2015}%
  \BibitemOpen
  \bibfield  {author} {\bibinfo {author} {\bibfnamefont {Z.}~\bibnamefont
  {Varga}}\ and\ \bibinfo {author} {\bibfnamefont {J.~W.}\ \bibnamefont
  {Swan}},\ }\bibfield  {title} {\enquote {\bibinfo {title} {Linear
  viscoelasticity of attractive colloidal dispersions},}\ }\href@noop {}
  {\bibfield  {journal} {\bibinfo  {journal} {Journal of Rheology}\ }\textbf
  {\bibinfo {volume} {59}},\ \bibinfo {pages} {1271--1298} (\bibinfo {year}
  {2015})}\BibitemShut {NoStop}%
\bibitem [{\citenamefont {Landrum}, \citenamefont {Russel},\ and\ \citenamefont
  {Zia}(2016)}]{Landrum:2016}%
  \BibitemOpen
  \bibfield  {author} {\bibinfo {author} {\bibfnamefont {B.~J.}\ \bibnamefont
  {Landrum}}, \bibinfo {author} {\bibfnamefont {W.~B.}\ \bibnamefont {Russel}},
  \ and\ \bibinfo {author} {\bibfnamefont {R.~N.}\ \bibnamefont {Zia}},\
  }\bibfield  {title} {\enquote {\bibinfo {title} {Delayed yield in colloidal
  gels: Creep, flow, and re-entrant solid regimes},}\ }\href@noop {} {\bibfield
   {journal} {\bibinfo  {journal} {Journal of Rheology}\ }\textbf {\bibinfo
  {volume} {60}},\ \bibinfo {pages} {783--807} (\bibinfo {year}
  {2016})}\BibitemShut {NoStop}%
\bibitem [{\citenamefont {Jamali}, \citenamefont {McKinley},\ and\
  \citenamefont {Armstrong}(2017)}]{Jamali:2017}%
  \BibitemOpen
  \bibfield  {author} {\bibinfo {author} {\bibfnamefont {S.}~\bibnamefont
  {Jamali}}, \bibinfo {author} {\bibfnamefont {G.~H.}\ \bibnamefont
  {McKinley}}, \ and\ \bibinfo {author} {\bibfnamefont {R.~C.}\ \bibnamefont
  {Armstrong}},\ }\bibfield  {title} {\enquote {\bibinfo {title}
  {Microstructural rearrangements and their rheological implications in a model
  thixotropic elastoviscoplastic fluid},}\ }\href@noop {} {\bibfield  {journal}
  {\bibinfo  {journal} {Phys. Rev. Lett.}\ }\textbf {\bibinfo {volume} {118}},\
  \bibinfo {pages} {048003} (\bibinfo {year} {2017})}\BibitemShut {NoStop}%
\bibitem [{\citenamefont {Bouzid}\ \emph {et~al.}(2017)\citenamefont {Bouzid},
  \citenamefont {Colombo}, \citenamefont {Barbosa},\ and\ \citenamefont {{Del
  Gado}}}]{Bouzid:2017}%
  \BibitemOpen
  \bibfield  {author} {\bibinfo {author} {\bibfnamefont {M.}~\bibnamefont
  {Bouzid}}, \bibinfo {author} {\bibfnamefont {J.}~\bibnamefont {Colombo}},
  \bibinfo {author} {\bibfnamefont {L.~V.}\ \bibnamefont {Barbosa}}, \ and\
  \bibinfo {author} {\bibfnamefont {E.}~\bibnamefont {{Del Gado}}},\ }\bibfield
   {title} {\enquote {\bibinfo {title} {Elastically driven intermittent
  microscopic dynamics in soft solids},}\ }\href@noop {} {\bibfield  {journal}
  {\bibinfo  {journal} {Nature Comm.}\ }\textbf {\bibinfo {volume} {8}},\
  \bibinfo {pages} {15846} (\bibinfo {year} {2017})}\BibitemShut {NoStop}%
\bibitem [{\citenamefont {Bouzid}\ and\ \citenamefont
  {Del~Gado}(2018)}]{bouzid2018langmuir}%
  \BibitemOpen
  \bibfield  {author} {\bibinfo {author} {\bibfnamefont {M.}~\bibnamefont
  {Bouzid}}\ and\ \bibinfo {author} {\bibfnamefont {E.}~\bibnamefont
  {Del~Gado}},\ }\bibfield  {title} {\enquote {\bibinfo {title} {Network
  topology in soft gels: Hardening and softening materials},}\ }\href {\doibase
  10.1021/acs.langmuir.7b02944} {\bibfield  {journal} {\bibinfo  {journal}
  {Langmuir}\ }\textbf {\bibinfo {volume} {34}},\ \bibinfo {pages} {773--781}
  (\bibinfo {year} {2018})}\BibitemShut {NoStop}%
\bibitem [{\citenamefont {Colombo}, \citenamefont {Widmer-Cooper},\ and\
  \citenamefont {{Del Gado}}(2013)}]{Colombo:2013}%
  \BibitemOpen
  \bibfield  {author} {\bibinfo {author} {\bibfnamefont {J.}~\bibnamefont
  {Colombo}}, \bibinfo {author} {\bibfnamefont {A.}~\bibnamefont
  {Widmer-Cooper}}, \ and\ \bibinfo {author} {\bibfnamefont {E.}~\bibnamefont
  {{Del Gado}}},\ }\bibfield  {title} {\enquote {\bibinfo {title} {Microscopic
  picture of cooperative processes in restructuring gel networks},}\
  }\href@noop {} {\bibfield  {journal} {\bibinfo  {journal} {Phys. Rev. Lett.}\
  }\textbf {\bibinfo {volume} {110}},\ \bibinfo {pages} {198301} (\bibinfo
  {year} {2013})}\BibitemShut {NoStop}%
\bibitem [{\citenamefont {Visscher}, \citenamefont {Mitchell},\ and\
  \citenamefont {Heyes}(1994)}]{Visscher1994}%
  \BibitemOpen
  \bibfield  {author} {\bibinfo {author} {\bibfnamefont {P.~B.}\ \bibnamefont
  {Visscher}}, \bibinfo {author} {\bibfnamefont {P.~J.}\ \bibnamefont
  {Mitchell}}, \ and\ \bibinfo {author} {\bibfnamefont {D.~M.}\ \bibnamefont
  {Heyes}},\ }\bibfield  {title} {\enquote {\bibinfo {title} {{Dynamic moduli
  of concentrated dispersions by Brownian dynamics}},}\ }\href@noop {}
  {\bibfield  {journal} {\bibinfo  {journal} {Journal of Rheology}\ }\textbf
  {\bibinfo {volume} {38}},\ \bibinfo {pages} {465--483} (\bibinfo {year}
  {1994})}\BibitemShut {NoStop}%
\bibitem [{\citenamefont {Au}\ and\ \citenamefont {Simmons}(2007)}]{Au2007}%
  \BibitemOpen
  \bibfield  {author} {\bibinfo {author} {\bibfnamefont {W.~W.~L.}\
  \bibnamefont {Au}}\ and\ \bibinfo {author} {\bibfnamefont {J.~A.}\
  \bibnamefont {Simmons}},\ }\bibfield  {title} {\enquote {\bibinfo {title}
  {{Echolocation in dolphins and bats}},}\ }\href@noop {} {\bibfield  {journal}
  {\bibinfo  {journal} {Physics Today}\ }\textbf {\bibinfo {volume} {60}},\
  \bibinfo {pages} {40--45} (\bibinfo {year} {2007})}\BibitemShut {NoStop}%
\bibitem [{\citenamefont {Geri}\ \emph {et~al.}(2018)\citenamefont {Geri},
  \citenamefont {Keshavarz}, \citenamefont {Divoux}, \citenamefont {Clasen},
  \citenamefont {Curtis},\ and\ \citenamefont {McKinley}}]{Geri:2018}%
  \BibitemOpen
  \bibfield  {author} {\bibinfo {author} {\bibfnamefont {M.}~\bibnamefont
  {Geri}}, \bibinfo {author} {\bibfnamefont {B.}~\bibnamefont {Keshavarz}},
  \bibinfo {author} {\bibfnamefont {T.}~\bibnamefont {Divoux}}, \bibinfo
  {author} {\bibfnamefont {C.}~\bibnamefont {Clasen}}, \bibinfo {author}
  {\bibfnamefont {D.~J.}\ \bibnamefont {Curtis}}, \ and\ \bibinfo {author}
  {\bibfnamefont {G.~H.}\ \bibnamefont {McKinley}},\ }\bibfield  {title}
  {\enquote {\bibinfo {title} {Time-resolved mechanical spectroscopy of soft
  materials via optimally windowed chirps},}\ }\href@noop {} {\bibfield
  {journal} {\bibinfo  {journal} {arXiv preprint arXiv:1804.03061}\ } (\bibinfo
  {year} {2018})}\BibitemShut {NoStop}%
\bibitem [{\citenamefont {Chasset}\ and\ \citenamefont
  {Thirion}(1965)}]{Chasset:1965}%
  \BibitemOpen
  \bibfield  {author} {\bibinfo {author} {\bibfnamefont {R.}~\bibnamefont
  {Chasset}}\ and\ \bibinfo {author} {\bibfnamefont {P.}~\bibnamefont
  {Thirion}},\ }\bibfield  {title} {\enquote {\bibinfo {title} {Viscoelastic
  relaxation of rubber vulcanizates between the glass transition and
  equilibrium},}\ }in\ \href@noop {} {\emph {\bibinfo {booktitle} {Proceedings
  of the Conference on Physics of Non-Crystalline Solids}}}\ (\bibinfo
  {publisher} {North-Holland Publ. Co., Amsterdam},\ \bibinfo {year} {1965})\
  pp.\ \bibinfo {pages} {345--357}\BibitemShut {NoStop}%
\bibitem [{\citenamefont {Chambon}\ \emph {et~al.}(1986)\citenamefont
  {Chambon}, \citenamefont {Petrovic}, \citenamefont {MacKnight},\ and\
  \citenamefont {Winter}}]{Chambon:1986}%
  \BibitemOpen
  \bibfield  {author} {\bibinfo {author} {\bibfnamefont {F.}~\bibnamefont
  {Chambon}}, \bibinfo {author} {\bibfnamefont {Z.~S.}\ \bibnamefont
  {Petrovic}}, \bibinfo {author} {\bibfnamefont {W.~J.}\ \bibnamefont
  {MacKnight}}, \ and\ \bibinfo {author} {\bibfnamefont {H.~H.}\ \bibnamefont
  {Winter}},\ }\bibfield  {title} {\enquote {\bibinfo {title} {Rheology of
  model polyurethanes at the gel point},}\ }\href@noop {} {\bibfield  {journal}
  {\bibinfo  {journal} {Macromolecules}\ }\textbf {\bibinfo {volume} {19}},\
  \bibinfo {pages} {2146--2149} (\bibinfo {year} {1986})}\BibitemShut {NoStop}%
\bibitem [{\citenamefont {Winter}\ and\ \citenamefont
  {Mours}(1997)}]{Winter:1997}%
  \BibitemOpen
  \bibfield  {author} {\bibinfo {author} {\bibfnamefont {H.~H.}\ \bibnamefont
  {Winter}}\ and\ \bibinfo {author} {\bibfnamefont {M.}~\bibnamefont {Mours}},\
  }\bibfield  {title} {\enquote {\bibinfo {title} {{Rheology of polymers near
  liquid-solid transitions}},}\ }\href@noop {} {\bibfield  {journal} {\bibinfo
  {journal} {Adv. Polym. Sci.}\ }\textbf {\bibinfo {volume} {134}},\ \bibinfo
  {pages} {165--234} (\bibinfo {year} {1997})}\BibitemShut {NoStop}%
\bibitem [{\citenamefont {Holt}, \citenamefont {Tripathi},\ and\ \citenamefont
  {Morgan}(2008)}]{Holt:2008}%
  \BibitemOpen
  \bibfield  {author} {\bibinfo {author} {\bibfnamefont {B.}~\bibnamefont
  {Holt}}, \bibinfo {author} {\bibfnamefont {A.}~\bibnamefont {Tripathi}}, \
  and\ \bibinfo {author} {\bibfnamefont {J.}~\bibnamefont {Morgan}},\
  }\bibfield  {title} {\enquote {\bibinfo {title} {Viscoelastic response of
  human skin to low magnitude physiologically relevant shear},}\ }\href@noop {}
  {\bibfield  {journal} {\bibinfo  {journal} {J Biomech.}\ }\textbf {\bibinfo
  {volume} {41}},\ \bibinfo {pages} {2689--2695} (\bibinfo {year}
  {2008})}\BibitemShut {NoStop}%
\bibitem [{\citenamefont {Nicolle}, \citenamefont {Vezin},\ and\ \citenamefont
  {J.-F.Palierne}(2010)}]{Nicolle:2010}%
  \BibitemOpen
  \bibfield  {author} {\bibinfo {author} {\bibfnamefont {S.}~\bibnamefont
  {Nicolle}}, \bibinfo {author} {\bibfnamefont {P.}~\bibnamefont {Vezin}}, \
  and\ \bibinfo {author} {\bibnamefont {J.-F.Palierne}},\ }\bibfield  {title}
  {\enquote {\bibinfo {title} {A strain-hardening bi-power law for the
  nonlinear behaviour of biological soft tissues},}\ }\href@noop {} {\bibfield
  {journal} {\bibinfo  {journal} {J. Biomech.}\ }\textbf {\bibinfo {volume}
  {43}},\ \bibinfo {pages} {927--932} (\bibinfo {year} {2010})}\BibitemShut
  {NoStop}%
\bibitem [{\citenamefont {Koos}\ and\ \citenamefont
  {Willenbacher}(2011)}]{Koos:2011}%
  \BibitemOpen
  \bibfield  {author} {\bibinfo {author} {\bibfnamefont {E.}~\bibnamefont
  {Koos}}\ and\ \bibinfo {author} {\bibfnamefont {N.}~\bibnamefont
  {Willenbacher}},\ }\bibfield  {title} {\enquote {\bibinfo {title} {Capillary
  forces in suspension rheology},}\ }\href@noop {} {\bibfield  {journal}
  {\bibinfo  {journal} {Science}\ }\textbf {\bibinfo {volume} {331}},\ \bibinfo
  {pages} {897--900} (\bibinfo {year} {2011})}\BibitemShut {NoStop}%
\bibitem [{\citenamefont {Curro}\ and\ \citenamefont
  {Pincus}(1983{\natexlab{a}})}]{Curro:1983}%
  \BibitemOpen
  \bibfield  {author} {\bibinfo {author} {\bibfnamefont {J.~G.}\ \bibnamefont
  {Curro}}\ and\ \bibinfo {author} {\bibfnamefont {P.}~\bibnamefont {Pincus}},\
  }\bibfield  {title} {\enquote {\bibinfo {title} {A theoretical basis for
  viscoelastic relaxation of elastomers in the long-time limit},}\ }\href@noop
  {} {\bibfield  {journal} {\bibinfo  {journal} {Macromolecules}\ }\textbf
  {\bibinfo {volume} {16}},\ \bibinfo {pages} {559--562} (\bibinfo {year}
  {1983}{\natexlab{a}})}\BibitemShut {NoStop}%
\bibitem [{\citenamefont {McKenna}\ and\ \citenamefont
  {Gaylord}(1988)}]{McKenna:1988}%
  \BibitemOpen
  \bibfield  {author} {\bibinfo {author} {\bibfnamefont {G.~B.}\ \bibnamefont
  {McKenna}}\ and\ \bibinfo {author} {\bibfnamefont {R.~J.}\ \bibnamefont
  {Gaylord}},\ }\bibfield  {title} {\enquote {\bibinfo {title} {Relaxation of
  crosslinked networks - theoretical-models and apparent power law behavior},}\
  }\href@noop {} {\bibfield  {journal} {\bibinfo  {journal} {Polymer}\ }\textbf
  {\bibinfo {volume} {29}},\ \bibinfo {pages} {2027--2032} (\bibinfo {year}
  {1988})}\BibitemShut {NoStop}%
\bibitem [{\citenamefont {Colombo}\ and\ \citenamefont
  {Del~Gado}(2014)}]{Colombo:2014b}%
  \BibitemOpen
  \bibfield  {author} {\bibinfo {author} {\bibfnamefont {J.}~\bibnamefont
  {Colombo}}\ and\ \bibinfo {author} {\bibfnamefont {E.}~\bibnamefont
  {Del~Gado}},\ }\bibfield  {title} {\enquote {\bibinfo {title} {Self-assembly
  and cooperative dynamics of a model colloidal gel network},}\ }\href@noop {}
  {\bibfield  {journal} {\bibinfo  {journal} {Soft Matter}\ }\textbf {\bibinfo
  {volume} {10}},\ \bibinfo {pages} {4003--4015} (\bibinfo {year}
  {2014})}\BibitemShut {NoStop}%
\bibitem [{\citenamefont {Derec}\ \emph {et~al.}(2003)\citenamefont {Derec},
  \citenamefont {Ducouret}, \citenamefont {Ajdari},\ and\ \citenamefont
  {Lequeux}}]{Derec:2003}%
  \BibitemOpen
  \bibfield  {author} {\bibinfo {author} {\bibfnamefont {C.}~\bibnamefont
  {Derec}}, \bibinfo {author} {\bibfnamefont {G.}~\bibnamefont {Ducouret}},
  \bibinfo {author} {\bibfnamefont {A.}~\bibnamefont {Ajdari}}, \ and\ \bibinfo
  {author} {\bibfnamefont {F.}~\bibnamefont {Lequeux}},\ }\bibfield  {title}
  {\enquote {\bibinfo {title} {Aging and nonlinear rheology in suspensions of
  polyethylene oxide--protected silica particles},}\ }\href@noop {} {\bibfield
  {journal} {\bibinfo  {journal} {Phys. Rev. E}\ }\textbf {\bibinfo {volume}
  {67}},\ \bibinfo {pages} {061403} (\bibinfo {year} {2003})}\BibitemShut
  {NoStop}%
\bibitem [{\citenamefont {Rajaram}\ and\ \citenamefont
  {Mohraz}(2011)}]{Rajaram:2011}%
  \BibitemOpen
  \bibfield  {author} {\bibinfo {author} {\bibfnamefont {B.}~\bibnamefont
  {Rajaram}}\ and\ \bibinfo {author} {\bibfnamefont {A.}~\bibnamefont
  {Mohraz}},\ }\bibfield  {title} {\enquote {\bibinfo {title} {Dynamics of
  shear-induced yielding and flow in dilute colloidal gels},}\ }\href@noop {}
  {\bibfield  {journal} {\bibinfo  {journal} {Physical Review E}\ }\textbf
  {\bibinfo {volume} {84}},\ \bibinfo {pages} {011405} (\bibinfo {year}
  {2011})}\BibitemShut {NoStop}%
\bibitem [{\citenamefont {Sprakel}\ \emph {et~al.}(2011)\citenamefont
  {Sprakel}, \citenamefont {Lindstr\"om}, \citenamefont {Kodger},\ and\
  \citenamefont {Weitz}}]{Sprakel:2011}%
  \BibitemOpen
  \bibfield  {author} {\bibinfo {author} {\bibfnamefont {J.}~\bibnamefont
  {Sprakel}}, \bibinfo {author} {\bibfnamefont {S.}~\bibnamefont
  {Lindstr\"om}}, \bibinfo {author} {\bibfnamefont {T.}~\bibnamefont {Kodger}},
  \ and\ \bibinfo {author} {\bibfnamefont {D.}~\bibnamefont {Weitz}},\
  }\bibfield  {title} {\enquote {\bibinfo {title} {Stress enhancement in the
  delayed yielding of colloidal gels},}\ }\href@noop {} {\bibfield  {journal}
  {\bibinfo  {journal} {Phys. Rev. Lett.}\ }\textbf {\bibinfo {volume} {106}},\
  \bibinfo {pages} {248303} (\bibinfo {year} {2011})}\BibitemShut {NoStop}%
\bibitem [{\citenamefont {Grenard}\ \emph {et~al.}(2014)\citenamefont
  {Grenard}, \citenamefont {Divoux}, \citenamefont {Taberlet},\ and\
  \citenamefont {Manneville}}]{Grenard:2014}%
  \BibitemOpen
  \bibfield  {author} {\bibinfo {author} {\bibfnamefont {V.}~\bibnamefont
  {Grenard}}, \bibinfo {author} {\bibfnamefont {T.}~\bibnamefont {Divoux}},
  \bibinfo {author} {\bibfnamefont {N.}~\bibnamefont {Taberlet}}, \ and\
  \bibinfo {author} {\bibfnamefont {S.}~\bibnamefont {Manneville}},\ }\bibfield
   {title} {\enquote {\bibinfo {title} {Timescales in creep and yielding of
  attractive gels},}\ }\href@noop {} {\bibfield  {journal} {\bibinfo  {journal}
  {Soft Matter}\ }\textbf {\bibinfo {volume} {10}},\ \bibinfo {pages}
  {1555--1571} (\bibinfo {year} {2014})}\BibitemShut {NoStop}%
\bibitem [{\citenamefont {Keshavarz}\ \emph {et~al.}(2017)\citenamefont
  {Keshavarz}, \citenamefont {Divoux}, \citenamefont {Manneville},\ and\
  \citenamefont {McKinley}}]{Keshavarz:2017}%
  \BibitemOpen
  \bibfield  {author} {\bibinfo {author} {\bibfnamefont {B.}~\bibnamefont
  {Keshavarz}}, \bibinfo {author} {\bibfnamefont {T.}~\bibnamefont {Divoux}},
  \bibinfo {author} {\bibfnamefont {S.}~\bibnamefont {Manneville}}, \ and\
  \bibinfo {author} {\bibfnamefont {G.~H.}\ \bibnamefont {McKinley}},\
  }\bibfield  {title} {\enquote {\bibinfo {title} {Nonlinear viscoelasticity
  and generalized failure criterion for polymer gels},}\ }\href@noop {}
  {\bibfield  {journal} {\bibinfo  {journal} {ACS Macro Letters}\ }\textbf
  {\bibinfo {volume} {6}},\ \bibinfo {pages} {663--667} (\bibinfo {year}
  {2017})}\BibitemShut {NoStop}%
\bibitem [{\citenamefont {Plimpton}(1995)}]{plimpton1995fast}%
  \BibitemOpen
  \bibfield  {author} {\bibinfo {author} {\bibfnamefont {S.}~\bibnamefont
  {Plimpton}},\ }\bibfield  {title} {\enquote {\bibinfo {title} {Fast parallel
  algorithms for short-range molecular dynamics},}\ }\href@noop {} {\bibfield
  {journal} {\bibinfo  {journal} {Journal of Computational Physics}\ }\textbf
  {\bibinfo {volume} {117}},\ \bibinfo {pages} {1--19} (\bibinfo {year}
  {1995})}\BibitemShut {NoStop}%
\bibitem [{\citenamefont {Lees}\ and\ \citenamefont
  {Edwards}(1972)}]{Lees:1972}%
  \BibitemOpen
  \bibfield  {author} {\bibinfo {author} {\bibfnamefont {A.}~\bibnamefont
  {Lees}}\ and\ \bibinfo {author} {\bibfnamefont {S.}~\bibnamefont {Edwards}},\
  }\bibfield  {title} {\enquote {\bibinfo {title} {The computer study of
  transport processes under extreme conditions},}\ }\href@noop {} {\bibfield
  {journal} {\bibinfo  {journal} {J. Phys. C: Solid State Phys.}\ }\textbf
  {\bibinfo {volume} {5(15)}},\ \bibinfo {pages} {1921--1929} (\bibinfo {year}
  {1972})}\BibitemShut {NoStop}%
\bibitem [{\citenamefont {Frenkel}\ and\ \citenamefont
  {Smit}(2001)}]{Frenkel:2001}%
  \BibitemOpen
  \bibfield  {author} {\bibinfo {author} {\bibfnamefont {D.}~\bibnamefont
  {Frenkel}}\ and\ \bibinfo {author} {\bibfnamefont {B.}~\bibnamefont {Smit}},\
  }\href@noop {} {\emph {\bibinfo {title} {Understanding Molecular
  Simulation}}}\ (\bibinfo  {publisher} {Elsevier},\ \bibinfo {year}
  {2001})\BibitemShut {NoStop}%
\bibitem [{\citenamefont {Irving}\ and\ \citenamefont
  {Kirkwood}(1950)}]{irving1950statistical}%
  \BibitemOpen
  \bibfield  {author} {\bibinfo {author} {\bibfnamefont {J.}~\bibnamefont
  {Irving}}\ and\ \bibinfo {author} {\bibfnamefont {J.~G.}\ \bibnamefont
  {Kirkwood}},\ }\bibfield  {title} {\enquote {\bibinfo {title} {The
  statistical mechanical theory of transport processes. iv. the equations of
  hydrodynamics},}\ }\href@noop {} {\bibfield  {journal} {\bibinfo  {journal}
  {The Journal of Chemical Physics}\ }\textbf {\bibinfo {volume} {18}},\
  \bibinfo {pages} {817--829} (\bibinfo {year} {1950})}\BibitemShut {NoStop}%
\bibitem [{\citenamefont {Macosko}(1994{\natexlab{a}})}]{Macosko:1994}%
  \BibitemOpen
  \bibfield  {author} {\bibinfo {author} {\bibfnamefont {C.}~\bibnamefont
  {Macosko}},\ }\href@noop {} {\emph {\bibinfo {title} {Rheology. Principles,
  measurements, and applications.}}}\ (\bibinfo  {publisher} {Wiley - VCH, New
  York},\ \bibinfo {year} {1994})\BibitemShut {NoStop}%
\bibitem [{\citenamefont {Pintelon}\ and\ \citenamefont
  {Schoukens}(2012)}]{Pintelon2012}%
  \BibitemOpen
  \bibfield  {author} {\bibinfo {author} {\bibfnamefont {R.}~\bibnamefont
  {Pintelon}}\ and\ \bibinfo {author} {\bibfnamefont {J.}~\bibnamefont
  {Schoukens}},\ }\href@noop {} {\emph {\bibinfo {title} {{System
  identification: a frequency domain approach}}}}\ (\bibinfo  {publisher} {John
  Wiley {\&} Sons},\ \bibinfo {year} {2012})\BibitemShut {NoStop}%
\bibitem [{\citenamefont {Winter}\ and\ \citenamefont
  {Chambon}(1986)}]{Winter:1986}%
  \BibitemOpen
  \bibfield  {author} {\bibinfo {author} {\bibfnamefont {H.~H.}\ \bibnamefont
  {Winter}}\ and\ \bibinfo {author} {\bibfnamefont {F.}~\bibnamefont
  {Chambon}},\ }\bibfield  {title} {\enquote {\bibinfo {title} {Analysis of
  linear viscoelasticity of a crosslinking polymer at the gel point},}\
  }\href@noop {} {\bibfield  {journal} {\bibinfo  {journal} {J. Rheol.}\
  }\textbf {\bibinfo {volume} {30}},\ \bibinfo {pages} {367--382} (\bibinfo
  {year} {1986})}\BibitemShut {NoStop}%
\bibitem [{\citenamefont {Mours}\ and\ \citenamefont
  {Winter}(1994)}]{Mours1994}%
  \BibitemOpen
  \bibfield  {author} {\bibinfo {author} {\bibfnamefont {M.}~\bibnamefont
  {Mours}}\ and\ \bibinfo {author} {\bibfnamefont {H.~H.}\ \bibnamefont
  {Winter}},\ }\bibfield  {title} {\enquote {\bibinfo {title} {{Time-resolved
  rheometry}},}\ }\href@noop {} {\bibfield  {journal} {\bibinfo  {journal}
  {Rheologica Acta}\ }\textbf {\bibinfo {volume} {33}},\ \bibinfo {pages}
  {385--397} (\bibinfo {year} {1994})}\BibitemShut {NoStop}%
\bibitem [{\citenamefont {Holly}\ \emph {et~al.}(1988)\citenamefont {Holly},
  \citenamefont {Venkataraman}, \citenamefont {Chambon},\ and\ \citenamefont
  {Winter}}]{Holly1988}%
  \BibitemOpen
  \bibfield  {author} {\bibinfo {author} {\bibfnamefont {E.~E.}\ \bibnamefont
  {Holly}}, \bibinfo {author} {\bibfnamefont {S.~K.}\ \bibnamefont
  {Venkataraman}}, \bibinfo {author} {\bibfnamefont {F.}~\bibnamefont
  {Chambon}}, \ and\ \bibinfo {author} {\bibfnamefont {H.~H.}\ \bibnamefont
  {Winter}},\ }\bibfield  {title} {\enquote {\bibinfo {title} {{Fourier
  transform mechanical spectroscopy of viscoelastic materials with transient
  structure}},}\ }\href@noop {} {\bibfield  {journal} {\bibinfo  {journal}
  {Journal of Non-Newtonian Fluid Mechanics}\ }\textbf {\bibinfo {volume}
  {27}},\ \bibinfo {pages} {17--26} (\bibinfo {year} {1988})}\BibitemShut
  {NoStop}%
\bibitem [{\citenamefont {Tang}\ \emph {et~al.}(2009)\citenamefont {Tang},
  \citenamefont {Saquing}, \citenamefont {Harding},\ and\ \citenamefont
  {Khan}}]{Tang2009}%
  \BibitemOpen
  \bibfield  {author} {\bibinfo {author} {\bibfnamefont {C.}~\bibnamefont
  {Tang}}, \bibinfo {author} {\bibfnamefont {C.~D.}\ \bibnamefont {Saquing}},
  \bibinfo {author} {\bibfnamefont {J.~R.}\ \bibnamefont {Harding}}, \ and\
  \bibinfo {author} {\bibfnamefont {S.~A.}\ \bibnamefont {Khan}},\ }\bibfield
  {title} {\enquote {\bibinfo {title} {{In situ cross-linking of electrospun
  poly (vinyl alcohol) nanofibers}},}\ }\href@noop {} {\bibfield  {journal}
  {\bibinfo  {journal} {Macromolecules}\ }\textbf {\bibinfo {volume} {43}},\
  \bibinfo {pages} {630--637} (\bibinfo {year} {2009})}\BibitemShut {NoStop}%
\bibitem [{\citenamefont {In}\ and\ \citenamefont
  {{Prud'homme}}(1993)}]{In1993}%
  \BibitemOpen
  \bibfield  {author} {\bibinfo {author} {\bibfnamefont {M.}~\bibnamefont
  {In}}\ and\ \bibinfo {author} {\bibfnamefont {R.~K.}\ \bibnamefont
  {{Prud'homme}}},\ }\bibfield  {title} {\enquote {\bibinfo {title} {{Fourier
  transform mechanical spectroscopy of the sol-gel transition in zirconium
  alkoxide ceramic gels}},}\ }\href@noop {} {\bibfield  {journal} {\bibinfo
  {journal} {Rheologica Acta}\ }\textbf {\bibinfo {volume} {32}},\ \bibinfo
  {pages} {556--565} (\bibinfo {year} {1993})}\BibitemShut {NoStop}%
\bibitem [{\citenamefont {Ross-Murphy}(1994)}]{Ross-Murphy1994}%
  \BibitemOpen
  \bibfield  {author} {\bibinfo {author} {\bibfnamefont {S.~B.}\ \bibnamefont
  {Ross-Murphy}},\ }\bibfield  {title} {\enquote {\bibinfo {title}
  {{Rheological characterization of polymer gels and networks}},}\ }\href@noop
  {} {\bibfield  {journal} {\bibinfo  {journal} {Polymer Gels and Networks}\
  }\textbf {\bibinfo {volume} {2}},\ \bibinfo {pages} {229--237} (\bibinfo
  {year} {1994})}\BibitemShut {NoStop}%
\bibitem [{\citenamefont {Pogodina}, \citenamefont {Winter},\ and\
  \citenamefont {Srinivas}(1999)}]{Pogodina1999}%
  \BibitemOpen
  \bibfield  {author} {\bibinfo {author} {\bibfnamefont {N.~V.}\ \bibnamefont
  {Pogodina}}, \bibinfo {author} {\bibfnamefont {H.~H.}\ \bibnamefont
  {Winter}}, \ and\ \bibinfo {author} {\bibfnamefont {S.}~\bibnamefont
  {Srinivas}},\ }\bibfield  {title} {\enquote {\bibinfo {title} {{Strain
  effects on physical gelation of crystallizing isotactic polypropylene}},}\
  }\href@noop {} {\bibfield  {journal} {\bibinfo  {journal} {Journal of Polymer
  Science Part B Polymer Physics}\ }\textbf {\bibinfo {volume} {37}},\ \bibinfo
  {pages} {3512--3519} (\bibinfo {year} {1999})}\BibitemShut {NoStop}%
\bibitem [{\citenamefont {Schwittay}, \citenamefont {Mours},\ and\
  \citenamefont {Winter}(1995)}]{Schwittay1995}%
  \BibitemOpen
  \bibfield  {author} {\bibinfo {author} {\bibfnamefont {C.}~\bibnamefont
  {Schwittay}}, \bibinfo {author} {\bibfnamefont {M.}~\bibnamefont {Mours}}, \
  and\ \bibinfo {author} {\bibfnamefont {H.~H.}\ \bibnamefont {Winter}},\
  }\bibfield  {title} {\enquote {\bibinfo {title} {{Rheological expression of
  physical gelation in polymers}},}\ }\href@noop {} {\bibfield  {journal}
  {\bibinfo  {journal} {Faraday Discussions}\ }\textbf {\bibinfo {volume}
  {101}},\ \bibinfo {pages} {93--104} (\bibinfo {year} {1995})}\BibitemShut
  {NoStop}%
\bibitem [{\citenamefont {Chiou}, \citenamefont {English},\ and\ \citenamefont
  {Khan}(1996)}]{Chiou1996}%
  \BibitemOpen
  \bibfield  {author} {\bibinfo {author} {\bibfnamefont {B.-S.}\ \bibnamefont
  {Chiou}}, \bibinfo {author} {\bibfnamefont {R.~J.}\ \bibnamefont {English}},
  \ and\ \bibinfo {author} {\bibfnamefont {S.~A.}\ \bibnamefont {Khan}},\
  }\bibfield  {title} {\enquote {\bibinfo {title} {{Rheology and
  Photo-Cross-Linking of Thiol-Ene Polymers}},}\ }\href@noop {} {\bibfield
  {journal} {\bibinfo  {journal} {Macromolecules}\ }\textbf {\bibinfo {volume}
  {29}},\ \bibinfo {pages} {5368--5374} (\bibinfo {year} {1996})}\BibitemShut
  {NoStop}%
\bibitem [{\citenamefont {Klauder}\ \emph {et~al.}(1960)\citenamefont
  {Klauder}, \citenamefont {Price}, \citenamefont {Darlington},\ and\
  \citenamefont {Albersheim}}]{Klauder1960}%
  \BibitemOpen
  \bibfield  {author} {\bibinfo {author} {\bibfnamefont {J.~R.}\ \bibnamefont
  {Klauder}}, \bibinfo {author} {\bibfnamefont {A.~C.}\ \bibnamefont {Price}},
  \bibinfo {author} {\bibfnamefont {S.}~\bibnamefont {Darlington}}, \ and\
  \bibinfo {author} {\bibfnamefont {W.~J.}\ \bibnamefont {Albersheim}},\
  }\bibfield  {title} {\enquote {\bibinfo {title} {{The theory and design of
  chirp radars}},}\ }\href@noop {} {\bibfield  {journal} {\bibinfo  {journal}
  {Bell Labs Technical Journal}\ }\textbf {\bibinfo {volume} {39}},\ \bibinfo
  {pages} {745--808} (\bibinfo {year} {1960})}\BibitemShut {NoStop}%
\bibitem [{\citenamefont {Farina}(2000)}]{Farina2000}%
  \BibitemOpen
  \bibfield  {author} {\bibinfo {author} {\bibfnamefont {A.}~\bibnamefont
  {Farina}},\ }\bibfield  {title} {\enquote {\bibinfo {title} {{Simultaneous
  measurement of impulse response and distortion with a swept-sine
  technique}},}\ }in\ \href@noop {} {\emph {\bibinfo {booktitle} {Audio
  Engineering Society Convention 108}}}\ (\bibinfo  {publisher} {Audio
  Engineering Society},\ \bibinfo {year} {2000})\BibitemShut {NoStop}%
\bibitem [{\citenamefont {Fausti}\ and\ \citenamefont
  {Farina}(2000)}]{Fausti2000}%
  \BibitemOpen
  \bibfield  {author} {\bibinfo {author} {\bibfnamefont {P.}~\bibnamefont
  {Fausti}}\ and\ \bibinfo {author} {\bibfnamefont {A.}~\bibnamefont
  {Farina}},\ }\bibfield  {title} {\enquote {\bibinfo {title} {{Acoustic
  measurements in opera houses: comparison between different techniques and
  equipment}},}\ }\href@noop {} {\bibfield  {journal} {\bibinfo  {journal}
  {Journal of Sound and Vibration}\ }\textbf {\bibinfo {volume} {232}},\
  \bibinfo {pages} {213--229} (\bibinfo {year} {2000})}\BibitemShut {NoStop}%
\bibitem [{\citenamefont {Ghiringhelli}\ \emph {et~al.}(2012)\citenamefont
  {Ghiringhelli}, \citenamefont {Roux}, \citenamefont {Bleses}, \citenamefont
  {Galliard},\ and\ \citenamefont {Caton}}]{Ghiringhelli:2012}%
  \BibitemOpen
  \bibfield  {author} {\bibinfo {author} {\bibfnamefont {E.}~\bibnamefont
  {Ghiringhelli}}, \bibinfo {author} {\bibfnamefont {D.}~\bibnamefont {Roux}},
  \bibinfo {author} {\bibfnamefont {D.}~\bibnamefont {Bleses}}, \bibinfo
  {author} {\bibfnamefont {H.}~\bibnamefont {Galliard}}, \ and\ \bibinfo
  {author} {\bibfnamefont {F.}~\bibnamefont {Caton}},\ }\bibfield  {title}
  {\enquote {\bibinfo {title} {Optimal fourier rheometry: Application to the
  gelation of an alginate},}\ }\href@noop {} {\bibfield  {journal} {\bibinfo
  {journal} {Rheol Acta}\ }\textbf {\bibinfo {volume} {51}},\ \bibinfo {pages}
  {413--420} (\bibinfo {year} {2012})}\BibitemShut {NoStop}%
\bibitem [{\citenamefont {Curtis}\ \emph {et~al.}(2015)\citenamefont {Curtis},
  \citenamefont {Holder}, \citenamefont {Badiei}, \citenamefont {Claypole},
  \citenamefont {Walters}, \citenamefont {Thomas}, \citenamefont {Barrow},
  \citenamefont {Deganello}, \citenamefont {Brown}, \citenamefont {Williams},\
  and\ \citenamefont {Hawkins}}]{Curtis:2015}%
  \BibitemOpen
  \bibfield  {author} {\bibinfo {author} {\bibfnamefont {D.}~\bibnamefont
  {Curtis}}, \bibinfo {author} {\bibfnamefont {A.}~\bibnamefont {Holder}},
  \bibinfo {author} {\bibfnamefont {N.}~\bibnamefont {Badiei}}, \bibinfo
  {author} {\bibfnamefont {J.}~\bibnamefont {Claypole}}, \bibinfo {author}
  {\bibfnamefont {M.}~\bibnamefont {Walters}}, \bibinfo {author} {\bibfnamefont
  {B.}~\bibnamefont {Thomas}}, \bibinfo {author} {\bibfnamefont
  {M.}~\bibnamefont {Barrow}}, \bibinfo {author} {\bibfnamefont
  {D.}~\bibnamefont {Deganello}}, \bibinfo {author} {\bibfnamefont
  {M.}~\bibnamefont {Brown}}, \bibinfo {author} {\bibfnamefont
  {P.}~\bibnamefont {Williams}}, \ and\ \bibinfo {author} {\bibfnamefont
  {K.}~\bibnamefont {Hawkins}},\ }\bibfield  {title} {\enquote {\bibinfo
  {title} {Validation of optimal fourier rheometry for rapidly gelling
  materials and its application in the study of collagen gelation},}\
  }\href@noop {} {\bibfield  {journal} {\bibinfo  {journal} {Journal of
  Non-Newtonian Fluid Mechanics}\ }\textbf {\bibinfo {volume} {222}},\ \bibinfo
  {pages} {253--259} (\bibinfo {year} {2015})}\BibitemShut {NoStop}%
\bibitem [{\citenamefont {Heyes}\ \emph {et~al.}(1994)\citenamefont {Heyes},
  \citenamefont {Mitchell}, \citenamefont {Visscher},\ and\ \citenamefont
  {Melrose}}]{Heyes1994}%
  \BibitemOpen
  \bibfield  {author} {\bibinfo {author} {\bibfnamefont {D.~M.}\ \bibnamefont
  {Heyes}}, \bibinfo {author} {\bibfnamefont {P.~J.}\ \bibnamefont {Mitchell}},
  \bibinfo {author} {\bibfnamefont {P.~B.}\ \bibnamefont {Visscher}}, \ and\
  \bibinfo {author} {\bibfnamefont {J.~R.}\ \bibnamefont {Melrose}},\
  }\bibfield  {title} {\enquote {\bibinfo {title} {{Brownian dynamics
  simulations of concentrated dispersions: viscoelasticity and near-Newtonian
  behaviour}},}\ }\href@noop {} {\bibfield  {journal} {\bibinfo  {journal}
  {Journal of the Chemical Society, Faraday Transactions}\ }\textbf {\bibinfo
  {volume} {90}},\ \bibinfo {pages} {1133--1141} (\bibinfo {year}
  {1994})}\BibitemShut {NoStop}%
\bibitem [{\citenamefont {Heyes}, \citenamefont {Mitchell},\ and\ \citenamefont
  {Visscher}(1994)}]{Heyes1994a}%
  \BibitemOpen
  \bibfield  {author} {\bibinfo {author} {\bibfnamefont {D.}~\bibnamefont
  {Heyes}}, \bibinfo {author} {\bibfnamefont {P.}~\bibnamefont {Mitchell}}, \
  and\ \bibinfo {author} {\bibfnamefont {P.}~\bibnamefont {Visscher}},\
  }\bibfield  {title} {\enquote {\bibinfo {title} {{Viscoelasticity and
  near-newtonian behaviour of concentrated dispersions by Brownian dynamics
  simulations}},}\ }\href@noop {} {\bibfield  {journal} {\bibinfo  {journal}
  {Trends in Colloid and Interface Science VIII}\ ,\ \bibinfo {pages}
  {179--182}} (\bibinfo {year} {1994})}\BibitemShut {NoStop}%
\bibitem [{\citenamefont {Blackman}\ and\ \citenamefont
  {Tukey}(1958)}]{Blackman1958}%
  \BibitemOpen
  \bibfield  {author} {\bibinfo {author} {\bibfnamefont {R.~B.}\ \bibnamefont
  {Blackman}}\ and\ \bibinfo {author} {\bibfnamefont {J.~W.}\ \bibnamefont
  {Tukey}},\ }\bibfield  {title} {\enquote {\bibinfo {title} {{The measurement
  of power spectra}},}\ }\href@noop {} {\bibfield  {journal} {\bibinfo
  {journal} {Bell Labs Technical Journal}\ }\textbf {\bibinfo {volume} {37}},\
  \bibinfo {pages} {185--282} (\bibinfo {year} {1958})}\BibitemShut {NoStop}%
\bibitem [{\citenamefont {Harris}(1978)}]{Harris1978}%
  \BibitemOpen
  \bibfield  {author} {\bibinfo {author} {\bibfnamefont {F.~J.}\ \bibnamefont
  {Harris}},\ }\bibfield  {title} {\enquote {\bibinfo {title} {{On the use of
  windows for harmonic analysis with the discrete Fourier transform}},}\
  }\href@noop {} {\bibfield  {journal} {\bibinfo  {journal} {Proceedings of the
  IEEE}\ }\textbf {\bibinfo {volume} {66}},\ \bibinfo {pages} {51--83}
  (\bibinfo {year} {1978})}\BibitemShut {NoStop}%
\bibitem [{\citenamefont {Tukey}(1967)}]{Tukey1967}%
  \BibitemOpen
  \bibfield  {author} {\bibinfo {author} {\bibfnamefont {J.~W.}\ \bibnamefont
  {Tukey}},\ }\bibfield  {title} {\enquote {\bibinfo {title} {{An introduction
  to the calculations of numerical spectrum analysis}},}\ }\href@noop {}
  {\bibfield  {journal} {\bibinfo  {journal} {Spectral analysis of time
  series}\ }\textbf {\bibinfo {volume} {25}} (\bibinfo {year}
  {1967})}\BibitemShut {NoStop}%
\bibitem [{\citenamefont {Walters}(1975)}]{Walters1975a}%
  \BibitemOpen
  \bibfield  {author} {\bibinfo {author} {\bibfnamefont {K.}~\bibnamefont
  {Walters}},\ }\href {https://books.google.com/books?id=kMZ1QgAACAAJ} {\emph
  {\bibinfo {title} {{Rheometry}}}}\ (\bibinfo  {publisher} {Chapman and
  Hall},\ \bibinfo {year} {1975})\BibitemShut {NoStop}%
\bibitem [{\citenamefont {Larson}(1999)}]{Larson:1999}%
  \BibitemOpen
  \bibfield  {author} {\bibinfo {author} {\bibfnamefont {R.~G.}\ \bibnamefont
  {Larson}},\ }\href@noop {} {\emph {\bibinfo {title} {The Structure and
  Rheology of Complex Fluids}}}\ (\bibinfo  {publisher} {Oxford University
  Press},\ \bibinfo {year} {1999})\BibitemShut {NoStop}%
\bibitem [{\citenamefont {Scott~Blair}\ and\ \citenamefont
  {Veinoglou}(1944)}]{Blair1944}%
  \BibitemOpen
  \bibfield  {author} {\bibinfo {author} {\bibfnamefont {G.~W.}\ \bibnamefont
  {Scott~Blair}}\ and\ \bibinfo {author} {\bibfnamefont {B.~C.}\ \bibnamefont
  {Veinoglou}},\ }\bibfield  {title} {\enquote {\bibinfo {title} {{A Study of
  the Firmness of Soft Materials Based on Nutting's Equation}},}\ }\href
  {\doibase 10.1088/0950-7671/21/9/301} {\bibfield  {journal} {\bibinfo
  {journal} {Journal of Scientific Instruments}\ }\textbf {\bibinfo {volume}
  {21}},\ \bibinfo {pages} {149--154} (\bibinfo {year} {1944})}\BibitemShut
  {NoStop}%
\bibitem [{\citenamefont {{Scott Blair}}(1944)}]{Blair1944a}%
  \BibitemOpen
  \bibfield  {author} {\bibinfo {author} {\bibfnamefont {G.~W.}\ \bibnamefont
  {{Scott Blair}}},\ }\bibfield  {title} {\enquote {\bibinfo {title}
  {{Analytical and Integrative Aspects of the Stress-Strain-Time Problem}},}\
  }\href {\doibase 10.1088/0950-7671/21/5/302} {\bibfield  {journal} {\bibinfo
  {journal} {Journal of Scientific Instruments}\ }\textbf {\bibinfo {volume}
  {21}},\ \bibinfo {pages} {80--84} (\bibinfo {year} {1944})}\BibitemShut
  {NoStop}%
\bibitem [{\citenamefont {Wagner}\ \emph {et~al.}(2017)\citenamefont {Wagner},
  \citenamefont {Barbati}, \citenamefont {Engmann}, \citenamefont {Burbidge},\
  and\ \citenamefont {McKinley}}]{Wagner:2017}%
  \BibitemOpen
  \bibfield  {author} {\bibinfo {author} {\bibfnamefont {C.~E.}\ \bibnamefont
  {Wagner}}, \bibinfo {author} {\bibfnamefont {A.~C.}\ \bibnamefont {Barbati}},
  \bibinfo {author} {\bibfnamefont {J.}~\bibnamefont {Engmann}}, \bibinfo
  {author} {\bibfnamefont {A.~S.}\ \bibnamefont {Burbidge}}, \ and\ \bibinfo
  {author} {\bibfnamefont {G.~H.}\ \bibnamefont {McKinley}},\ }\bibfield
  {title} {\enquote {\bibinfo {title} {Quantifying the consistency and rheology
  of liquid foods using fractional calculus},}\ }\href@noop {} {\bibfield
  {journal} {\bibinfo  {journal} {Food Hydrocolloids}\ }\textbf {\bibinfo
  {volume} {69}},\ \bibinfo {pages} {242--254} (\bibinfo {year}
  {2017})}\BibitemShut {NoStop}%
\bibitem [{\citenamefont {Podlubny}(1998)}]{Podlubny1998}%
  \BibitemOpen
  \bibfield  {author} {\bibinfo {author} {\bibfnamefont {I.}~\bibnamefont
  {Podlubny}},\ }\href@noop {} {\emph {\bibinfo {title} {Fractional
  differential equations: an introduction to fractional derivatives, fractional
  differential equations, to methods of their solution and some of their
  applications}}},\ Vol.\ \bibinfo {volume} {198}\ (\bibinfo  {publisher}
  {Academic press},\ \bibinfo {year} {1998})\BibitemShut {NoStop}%
\bibitem [{\citenamefont {Curro}\ and\ \citenamefont
  {Pincus}(1983{\natexlab{b}})}]{Curro1983}%
  \BibitemOpen
  \bibfield  {author} {\bibinfo {author} {\bibfnamefont {J.~G.}\ \bibnamefont
  {Curro}}\ and\ \bibinfo {author} {\bibfnamefont {P.}~\bibnamefont {Pincus}},\
  }\bibfield  {title} {\enquote {\bibinfo {title} {{A theoretical basis for
  viscoelastic relaxation of elastomers in the long-time limit}},}\ }\href@noop
  {} {\bibfield  {journal} {\bibinfo  {journal} {Macromolecules}\ }\textbf
  {\bibinfo {volume} {16}},\ \bibinfo {pages} {559--562} (\bibinfo {year}
  {1983}{\natexlab{b}})}\BibitemShut {NoStop}%
\bibitem [{\citenamefont {Schiessel}\ \emph {et~al.}(1995)\citenamefont
  {Schiessel}, \citenamefont {Metzler}, \citenamefont {Blumen},\ and\
  \citenamefont {Nonnenmacher}}]{Schiessel1995a}%
  \BibitemOpen
  \bibfield  {author} {\bibinfo {author} {\bibfnamefont {H.}~\bibnamefont
  {Schiessel}}, \bibinfo {author} {\bibfnamefont {R.}~\bibnamefont {Metzler}},
  \bibinfo {author} {\bibfnamefont {A.}~\bibnamefont {Blumen}}, \ and\ \bibinfo
  {author} {\bibfnamefont {T.~F.}\ \bibnamefont {Nonnenmacher}},\ }\bibfield
  {title} {\enquote {\bibinfo {title} {{Generalized viscoelastic models: Their
  fractional equations with solutions}},}\ }\href {\doibase
  10.1088/0305-4470/28/23/012} {\bibfield  {journal} {\bibinfo  {journal}
  {Journal of Physics A-Mathematical and General}\ }\textbf {\bibinfo {volume}
  {28}},\ \bibinfo {pages} {6567--6584} (\bibinfo {year} {1995})}\BibitemShut
  {NoStop}%
\bibitem [{\citenamefont {Bagley}\ and\ \citenamefont
  {Torvik}(1983)}]{Bagley:1983}%
  \BibitemOpen
  \bibfield  {author} {\bibinfo {author} {\bibfnamefont {R.~L.}\ \bibnamefont
  {Bagley}}\ and\ \bibinfo {author} {\bibfnamefont {P.~J.}\ \bibnamefont
  {Torvik}},\ }\bibfield  {title} {\enquote {\bibinfo {title} {A theoretical
  basis for the application of fractional calculus to viscoelasticity},}\
  }\href@noop {} {\bibfield  {journal} {\bibinfo  {journal} {Journal of
  Rheology}\ }\textbf {\bibinfo {volume} {27}},\ \bibinfo {pages} {201--210}
  (\bibinfo {year} {1983})}\BibitemShut {NoStop}%
\bibitem [{\citenamefont {Wharmby}\ and\ \citenamefont
  {Bagley}(2013)}]{Wharmby:2013}%
  \BibitemOpen
  \bibfield  {author} {\bibinfo {author} {\bibfnamefont {A.~W.}\ \bibnamefont
  {Wharmby}}\ and\ \bibinfo {author} {\bibfnamefont {R.~L.}\ \bibnamefont
  {Bagley}},\ }\bibfield  {title} {\enquote {\bibinfo {title} {Generalization
  of a theoretical basis for the application of fractional calculus to
  viscoelasticity},}\ }\href@noop {} {\bibfield  {journal} {\bibinfo  {journal}
  {J. Rheol.}\ }\textbf {\bibinfo {volume} {57}},\ \bibinfo {pages}
  {1429--1440} (\bibinfo {year} {2013})}\BibitemShut {NoStop}%
\bibitem [{\citenamefont {Friedrich}, \citenamefont {Schiessel},\ and\
  \citenamefont {Blumen}(1999)}]{Friedrich:1999}%
  \BibitemOpen
  \bibfield  {author} {\bibinfo {author} {\bibfnamefont {C.}~\bibnamefont
  {Friedrich}}, \bibinfo {author} {\bibfnamefont {H.}~\bibnamefont
  {Schiessel}}, \ and\ \bibinfo {author} {\bibfnamefont {A.}~\bibnamefont
  {Blumen}},\ }\enquote {\bibinfo {title} {Advances in the flow and rheology of
  non-newtonian fluids},}\ \ (\bibinfo {year} {1999})\ Chap.\ \bibinfo
  {chapter} {Constitutive Behavior Modeling and Fractional Derivatives}, pp.\
  \bibinfo {pages} {429--466}\BibitemShut {NoStop}%
\bibitem [{\citenamefont {Ng}\ and\ \citenamefont {McKinley}(2008)}]{Ng:2008}%
  \BibitemOpen
  \bibfield  {author} {\bibinfo {author} {\bibfnamefont {T.~S.~K.}\
  \bibnamefont {Ng}}\ and\ \bibinfo {author} {\bibfnamefont {G.~H.}\
  \bibnamefont {McKinley}},\ }\bibfield  {title} {\enquote {\bibinfo {title}
  {Power law gels at finite strains: The nonlinear rheology of gluten gels},}\
  }\href@noop {} {\bibfield  {journal} {\bibinfo  {journal} {Journal of
  Rheology}\ } (\bibinfo {year} {2008})}\BibitemShut {NoStop}%
\bibitem [{\citenamefont {Shih}\ \emph {et~al.}(1990)\citenamefont {Shih},
  \citenamefont {Shih}, \citenamefont {Kum}, \citenamefont {Liu},\ and\
  \citenamefont {Aksay}}]{Shih:1990}%
  \BibitemOpen
  \bibfield  {author} {\bibinfo {author} {\bibfnamefont {W.-H.}\ \bibnamefont
  {Shih}}, \bibinfo {author} {\bibfnamefont {W.~Y.}\ \bibnamefont {Shih}},
  \bibinfo {author} {\bibfnamefont {S.-I.}\ \bibnamefont {Kum}}, \bibinfo
  {author} {\bibfnamefont {J.}~\bibnamefont {Liu}}, \ and\ \bibinfo {author}
  {\bibfnamefont {I.~A.}\ \bibnamefont {Aksay}},\ }\bibfield  {title} {\enquote
  {\bibinfo {title} {Scaling behavior of the elastic properties of colloidal
  gels},}\ }\href@noop {} {\bibfield  {journal} {\bibinfo  {journal} {Phys.
  Rev. A}\ }\textbf {\bibinfo {volume} {42}},\ \bibinfo {pages} {4772--4779}
  (\bibinfo {year} {1990})}\BibitemShut {NoStop}%
\bibitem [{\citenamefont {Bremer}, \citenamefont {{van Vliet}},\ and\
  \citenamefont {Walstra}(1989)}]{Bremer:1989}%
  \BibitemOpen
  \bibfield  {author} {\bibinfo {author} {\bibfnamefont {L.}~\bibnamefont
  {Bremer}}, \bibinfo {author} {\bibfnamefont {T.}~\bibnamefont {{van Vliet}}},
  \ and\ \bibinfo {author} {\bibfnamefont {P.}~\bibnamefont {Walstra}},\
  }\bibfield  {title} {\enquote {\bibinfo {title} {Theoretical and experimental
  study of the fractal nature of the structure of casein gels},}\ }\href@noop
  {} {\bibfield  {journal} {\bibinfo  {journal} {J. Chem. Soc. Faraday Trans.
  I}\ }\textbf {\bibinfo {volume} {85}},\ \bibinfo {pages} {3359--3372}
  (\bibinfo {year} {1989})}\BibitemShut {NoStop}%
\bibitem [{\citenamefont {Curtis}\ \emph {et~al.}(2013)\citenamefont {Curtis},
  \citenamefont {Williams}, \citenamefont {Badiei}, \citenamefont {Campbell},
  \citenamefont {Hawkins}, \citenamefont {Evans},\ and\ \citenamefont
  {Brownd}}]{Curtis:2013}%
  \BibitemOpen
  \bibfield  {author} {\bibinfo {author} {\bibfnamefont {D.~J.}\ \bibnamefont
  {Curtis}}, \bibinfo {author} {\bibfnamefont {P.~R.}\ \bibnamefont
  {Williams}}, \bibinfo {author} {\bibfnamefont {N.}~\bibnamefont {Badiei}},
  \bibinfo {author} {\bibfnamefont {A.~I.}\ \bibnamefont {Campbell}}, \bibinfo
  {author} {\bibfnamefont {K.}~\bibnamefont {Hawkins}}, \bibinfo {author}
  {\bibfnamefont {P.~A.}\ \bibnamefont {Evans}}, \ and\ \bibinfo {author}
  {\bibfnamefont {M.~R.}\ \bibnamefont {Brownd}},\ }\bibfield  {title}
  {\enquote {\bibinfo {title} {A study of microstructural templating in
  fibrin–thrombin gel networks by spectral and viscoelastic analysis},}\
  }\href@noop {} {\bibfield  {journal} {\bibinfo  {journal} {Soft Matter}\
  }\textbf {\bibinfo {volume} {9}},\ \bibinfo {pages} {4883--4889} (\bibinfo
  {year} {2013})}\BibitemShut {NoStop}%
\bibitem [{\citenamefont {Hung}, \citenamefont {Jeng},\ and\ \citenamefont
  {Hsu}(2015)}]{Hung:2015}%
  \BibitemOpen
  \bibfield  {author} {\bibinfo {author} {\bibfnamefont {K.-C.}\ \bibnamefont
  {Hung}}, \bibinfo {author} {\bibfnamefont {U.-S.}\ \bibnamefont {Jeng}}, \
  and\ \bibinfo {author} {\bibfnamefont {S.-H.}\ \bibnamefont {Hsu}},\
  }\bibfield  {title} {\enquote {\bibinfo {title} {Fractal structure of
  hydrogels modulates stem cell behavior},}\ }\href@noop {} {\bibfield
  {journal} {\bibinfo  {journal} {ACS macro Letters}\ }\textbf {\bibinfo
  {volume} {4}},\ \bibinfo {pages} {1056--1061} (\bibinfo {year}
  {2015})}\BibitemShut {NoStop}%
\bibitem [{\citenamefont {Macosko}(1994{\natexlab{b}})}]{Macosko1994a}%
  \BibitemOpen
  \bibfield  {author} {\bibinfo {author} {\bibfnamefont {C.~W.}\ \bibnamefont
  {Macosko}},\ }\href
  {http://books.google.com/books/about/Rheology.html?id=Kai7QgAACAAJ{\&}pgis=1}
  {\emph {\bibinfo {title} {{Rheology: Principles, Measurements, and
  Applications}}}}\ (\bibinfo  {publisher} {Wiley},\ \bibinfo {year} {1994})\
  p.\ \bibinfo {pages} {568}\BibitemShut {NoStop}%
\end{thebibliography}%

\begin{acknowledgments}
This work was funded by the MIT-France seed fund and by the CNRS PICS-USA scheme (\#36939). BK acknowledges financial support from Axalta Coating Systems. MB and EDG thank the Impact Program of the Georgetown Environmental Initiative and Georgetown University for support.
\end{acknowledgments}

\section*{Appendix}
\label{appendix1}


\subsection{Varying protocol and total inertia}

\begin{figure}[h!]
\includegraphics[width=0.9\linewidth]{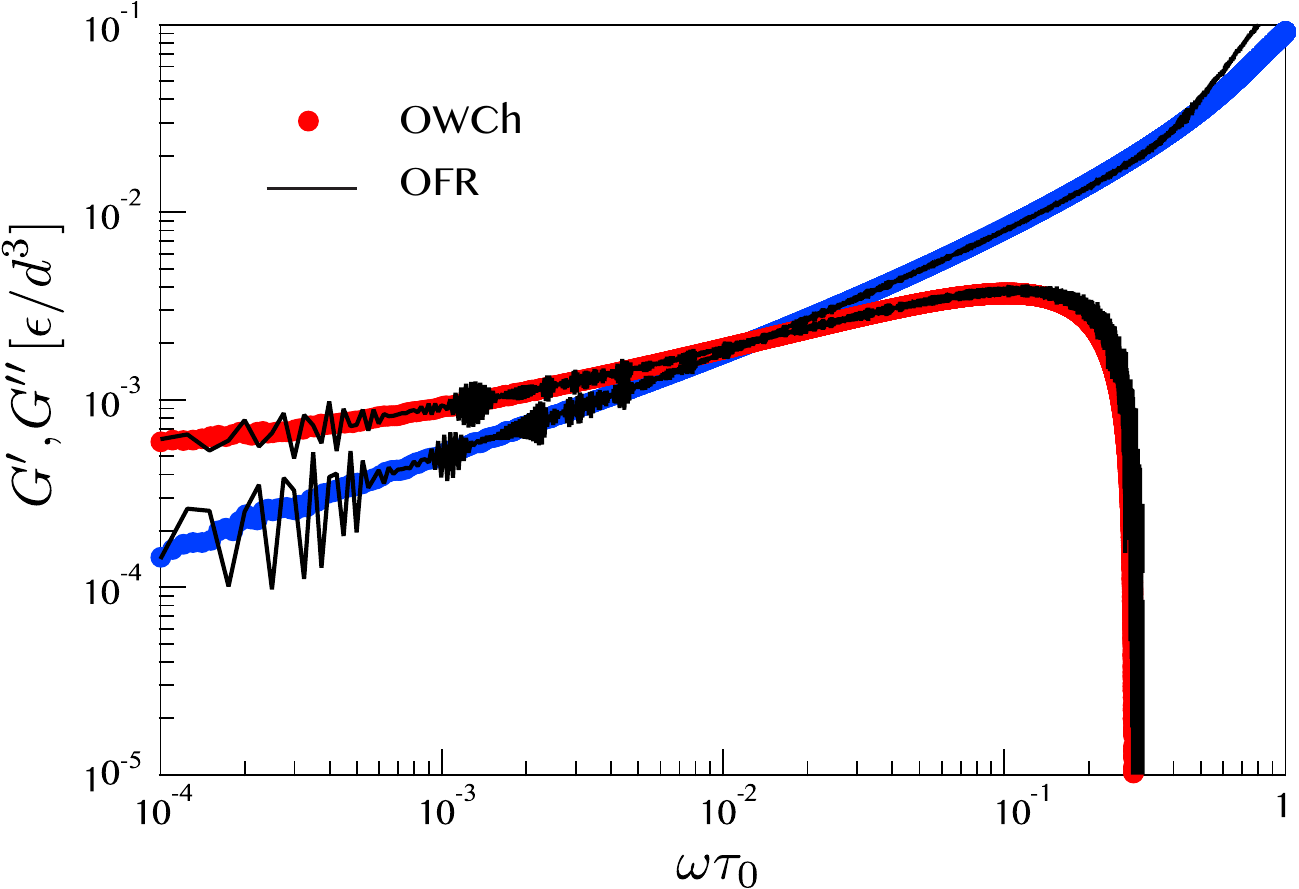}
\caption{Storage and loss moduli $G'$ and $G''$ as a function of the frequency $\omega$. Comparison between the spectrum computed with the OFR protocol ($-$) and the OWCh protocol ($\bullet$). The latter data set is displayed in Fig.~\ref{Fig.3}. The OWCh protocol significantly reduces the leakage error in the low frequency limit. 
\label{Fig.9}}
\end{figure}

Fig.~\ref{Fig.9} compares the frequency spectrum of the model particulate gel, computed with the Optimal Fourier Rheometry (OFR) protocol and the Optimally Windowed Chirped (OWCh) protocol. The OWCh leads to a more accurate picture of the frequency spectrum, especially in the limit of low frequency. See Ref. \cite{Geri:2018} for an extended comparison. 
\begin{figure}[!h]
\includegraphics[width=\linewidth]{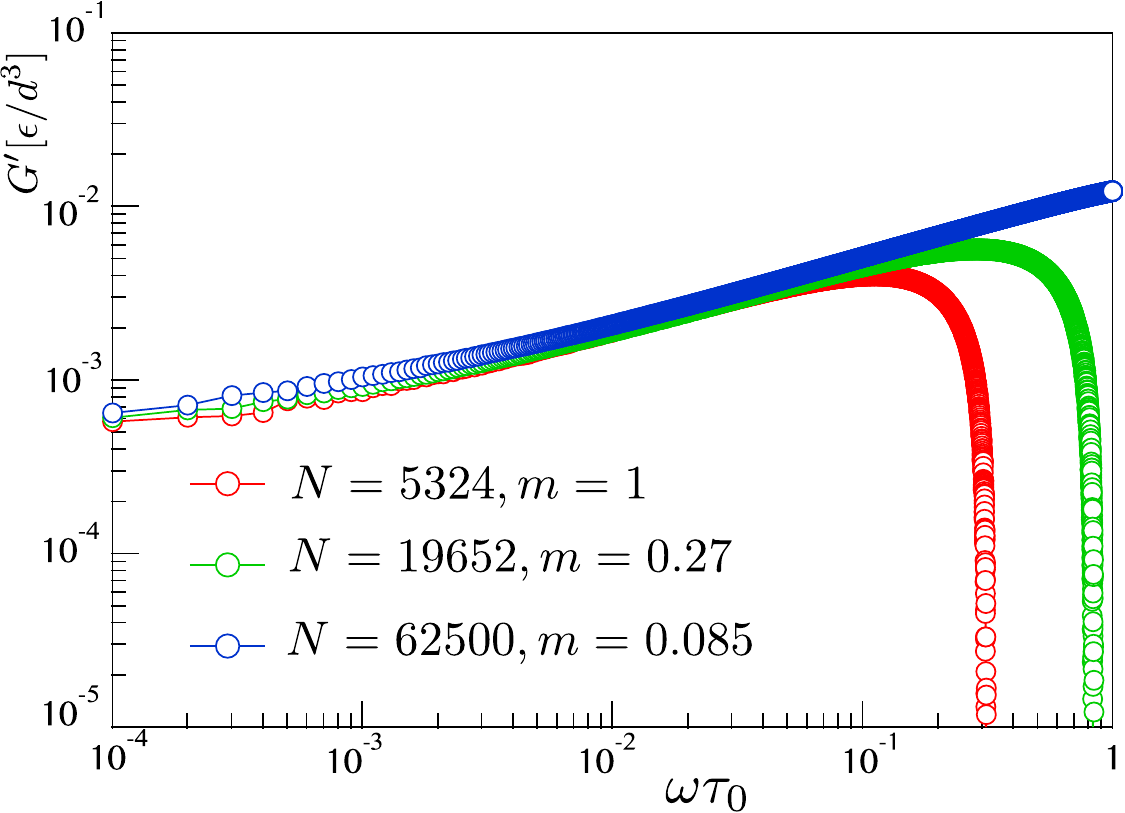}
\caption{Storage modulus $G'$ as a function of the frequency $\omega$ keeping $\bar{M}=N\times m=5324$ constant by varying the total number of particles $N$ and the particle mass $m$. Other parameters are set to $\phi\sim7.3\%$ and $\eta_f=0.5$.  
\label{Fig.10}}
\end{figure}
Fig.~\ref{Fig.10} illustrates the effect of varying the individual mass $m$ of particles on the resulting frequency dependence of $G'$ for a gel of constant total mass $\bar{M}=N\times m$, where $N$ denotes the total number of particles. Decreasing the particle mass does not affect the overall shape of the curve nor the power-law dependence of $G'(\omega)$ but leads to a progressive shift towards larger frequency of the position $\omega_c$ corresponding to the maximum of the elastic modulus.  

\begin{figure}[!b]
\includegraphics[trim = 10mm 05mm 20mm 20mm, clip,width=\columnwidth]{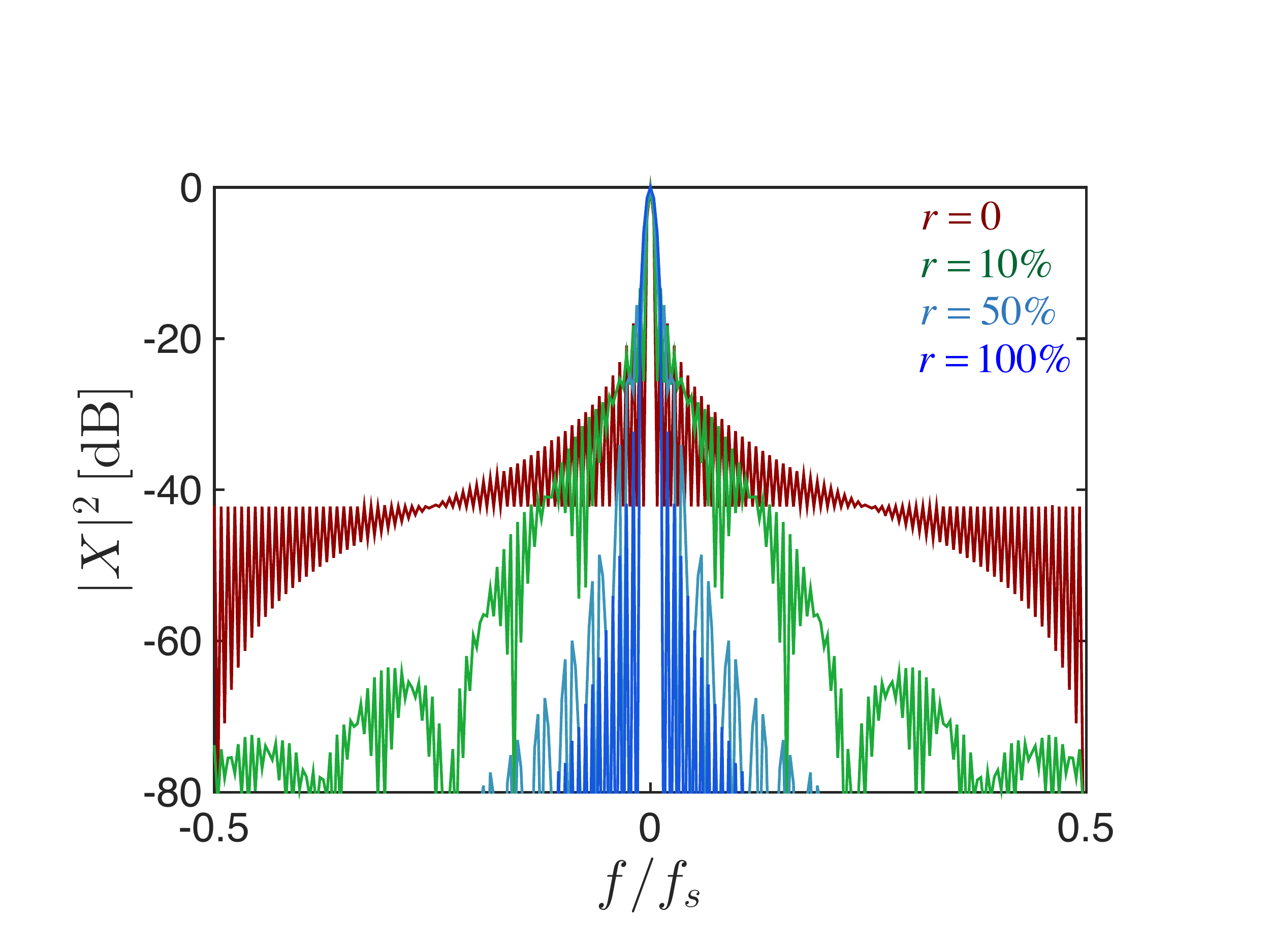}
\caption{The power spectrum of different symmetric Tukey cosine tapered windows. Different colors represent different values of the tapering parameter $r$. The $y$ axis is the square of the magnitude of the Fourier transform vector ($X(\omega)=\mathscr{F}\left\{w(t)\right\}$) and the $x$ axis is the frequency in the Fourier domain $f=\omega/2\pi$ normalized by the value of the sampling frequency $f_s$. \label{powerwindow}}
\end{figure}

\subsection{Power Spectrum of the Window Function}
\label{appendix2}
Fig.~\ref{powerwindow} shows the power spectrum of a series of amplitude envelopes that belong to the family of symmetric Tukey cosine-tapered windows with different values of tapering parameter $r$. These windows are the symmetric case of the one-sided tapered function that is used in this study (equation \ref{window}) with $r$ being equivalent to $b$. For the rectangular window, with $r=0$ being equivalent to the case of no amplitude modulation, the power spectrum has a peak at zero frequency but also has other local peaks at other frequencies. When we apply this window to the input strain signal, the corresponding Fourier transform will be the convolution of the individual Fourier transforms of the window and the strain signal. Thus, due to the inclusive nature of the convolution integral, the Fourier transform of the input signal will be affected by the contributions from all frequencies in the power spectrum of the window. This leads to the known spectral leakage error and in order to avoid it a window with narrower power spectrum is desired. It is evident that with increasing the tapering parameter the power spectrum of the window becomes narrower and the contributions of non-zero frequencies decay more rapidly. However, if we use a very high tapering parameter the power of the input signal is attenuated and may become comparable to the existing noise level. This explains why in our simulation we have used a moderate value for the tapering parameter $r=b=0.1$.

\subsection{Analogy with the Mechanical System and the Bode Plot}
\label{appendix3}
The frequency response of the studied gel can be understood by studying the vibrational response of its corresponding mechanical toy model. This simple analogy is often used in rheological measurements for understanding the effect of inertia in the system. A relevant example is the measurement of frequency response in stress-controlled rheometers when the inertia of the oscillating geometry is of significance \cite{Walters1975a,Macosko1994a}. As discussed by Walters \cite{Walters1975a} the inertia of the geometry, similar to the particle mass in our simulations, plays the role of a vibrating mass in the mechanical model and the elastic and viscous moduli of the material are analogous to the spring and dash-pot elements respectively. If we excite this system by an external force (around certain frequencies), it is known that  resonance can happen, which translates into an amplified level of oscillation and a sudden sign change for the phase angle. 
\begin{figure}[!h]
\includegraphics[trim = 30mm 0mm 40mm 0mm, clip,width=\columnwidth]{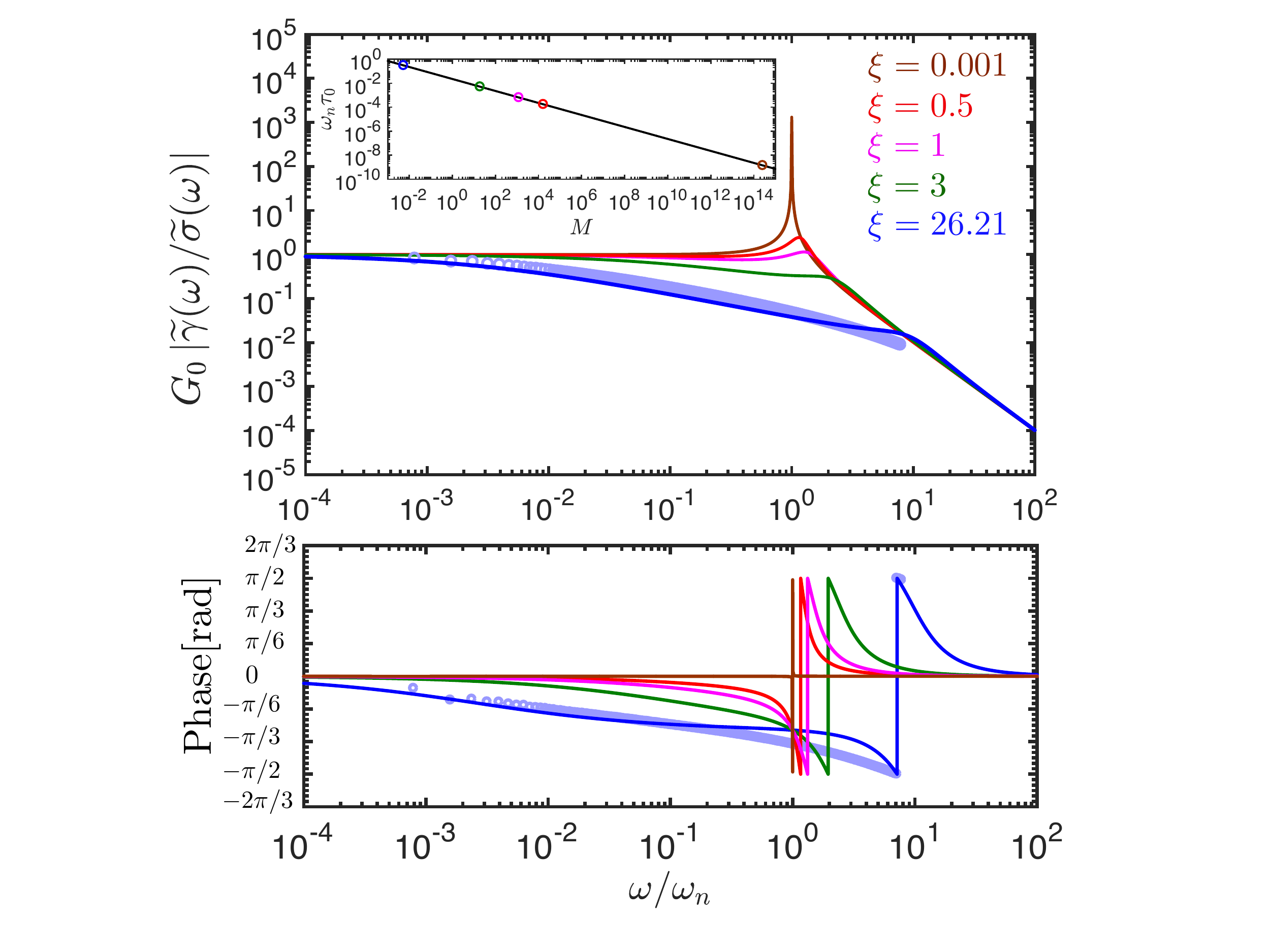}
\caption{Bode plots illustrating inertial resonance of a damped fractional viscoelastic system described by the Fractional Kelvin-Voigt Model with inertial mass $M$, and damping  $\xi\equiv\mathbb{V}/\sqrt{M^\alpha G_0^{2-\alpha}}$. Continuous lines of different colors represent different values of the damping ratio $\xi$ and the blue circles correspond to numerical simulations for a colloidal gels with $N=19652$ particles and of $m=0.25$. Inset: resonance frequency $\omega \tau_0$ vs the inertial mass $M$. Again the blue circle corresponds to a numerical simulation for a model colloidal gel with $N=19652$ particles and of $m=0.25$.}\label{bode}
\end{figure}
Resonance often indicates that the measured elastic modulus reported by the rheometer deviates from the intrinsic elastic modulus of the material and is dominated by the inertia of the oscillating mass. A very similar scenario happens in our simulations but it is due to the presence of particle inertia. In our model gels, one can think of the stress as the equivalent to the excitation force in the vibrating system. Similarly, the strain signal is equivalent to the amplitude response of the system. A common method for studying the response of a vibrating system is through construction of the corresponding Bode plots in the Fourier/frequency domain. In a typical Bode plot, we study the frequency behavior of the response function (normalized output deformation divided by the input excitation signal) $G_0\tilde{\gamma}(\omega)/\tilde{\sigma}(\omega)$ in terms of its corresponding magnitude and phase angle.

Figure \ref{bode} shows a series of Bode plots for the FKVM mechanical system that is discussed in the main text [see Fig.~\ref{Fig.1}(a)]. Different colors represent different values of the damping ratio $\xi\equiv\mathbb{V}/\sqrt{M^\alpha G_0^{2-\alpha}}$. In the simulations, we can change the damping ratio and vary the particle mass $m$. All the other parameters are kept constant. By analogy with the mechanical toy model, where the natural resonance frequency is $\omega_n=\sqrt{G_0/M}$,  the natural frequency of the numerical model also decreases with increasing the particle mass $m$. The inset in Figure~\ref{bode} shows that the resonance frequency diverges to infinity for zero $M$ (and zero particle mass). That is, in the limit of massless particles or systems with negligible inertia the onset of the resonance phenomena is shifted to such high frequencies that the whole phenomena of resonance may not be detectable in the range of studied frequencies in a typical experimental/numerical analysis. On the other hand, one can also observe that the resonance has two major effects on the measured response function of the material. First, as the Bode plot for the magnitude (top sub-plot in Figure~\ref{bode}) suggests, the emergence of a peak in the amplitude of the response function emphasizes the idea of amplified vibration due to inertial resonance. Second, the phase plot (bottom sub-plot in Figure \ref{bode}) clearly shows that as the system passes through the resonance there is a sign change in the phase angle ($\arctan(-G''/G')$) which is due to the fact that the
in phase contribution to the signal decrease from positive to negative values as the system passes through resonance in the frequency domain. Our numerical simulations for a colloidal gels with $N=19652$ particles and of $m=0.25$ (blue circles) show a very similar trend to the mechanical model. This again emphasizes the fact that in experimental/numerical systems for which inertia is included one should expect the onset of inertial resonance at a certain frequency.

Finally it is interesting to note that, in spite of the analogy and similarities between the numerical model composed of attractive particles with inertia and the case of experiments in which inertia is present due to the rheometer geometry, our study elucidates the following difference. In the case of the numerical model, for which the inertia is a property of the gel and arises due to particle inertia, the critical frequency $\omega_c$ at resonance depends on the individual particle mass $m$ and on the viscous damping $\eta_{f}$, but not on the total mass $\bar{M}$ (see Figs.\ref{Fig.8}, \ref{Fig.5} and Fig.\ref{Fig.10}).  In contrast, when the inertia is due to the fluid sample in the rheometer and/or the moment of inertia of the rheometer fixture, the resonance will depend on the sample size, i.e. vary with the total mass of the sample and hence with its volume (for a fixed volume fraction and type of particles). Such observations, combined with a suitable lumped parameter model of the form we outline here, can help identify (and at least partially correct for) the source of inertial effects that may be contaminating rheological measurements in situations where it is not immediately obvious how to distinguish different contributions.

\end{document}